\documentclass[twocolumn, notitlepage, secnumarabic,amssymb, aps, prl, figure, superscriptaddress, floatfix] {revtex4-1}
\usepackage{graphicx}
\usepackage{amsmath}
\usepackage{subfigure}
\usepackage{braket}
\usepackage{ulem}
\usepackage{hyperref}
\hypersetup{
    colorlinks=true,
    linkcolor=blue,
    filecolor=red,      
    urlcolor=blue,
    citecolor=magenta
}
\usepackage{soul}
\newcommand{\ctext}[3][RGB]{%
  \begingroup
  \definecolor{hlcolor}{#1}{#2}\sethlcolor{hlcolor}%
  \hl{#3}%
  \endgroup
}

\usepackage{ragged2e}
\usepackage{caption}
\DeclareCaptionJustification{just}{\RaggedRight}
\DeclareCaptionLabelSeparator{naturestyle}{\textbf{.} }
\DeclareCaptionFormat{naturestyle}{\textbf{#1}\textbf{#2}#3\par}
\DeclareCaptionLabelFormat{naturestyle}{\textbf{FIG. }\textbf{#2}}
\usepackage[usenames, dvipsnames]{color}
\definecolor{mypink}{RGB}{219, 48, 122}

\begin{document}

\title{Prediction of a supersolid phase in high-pressure deuterium}

\author{Chang Woo Myung}
\affiliation{Yusuf Hamied Department of Chemistry, University of Cambridge, Lensﬁeld Road, Cambridge, CB2 1EW, United Kingdom}

\author{Barak Hirshberg}
\email{hirshb@tauex.tau.ac.il}
\affiliation{School of Chemistry, Tel Aviv University, Tel Aviv 6997801, Israel}

\author{Michele Parrinello}
\email{michele.parrinello@iit.it}
\affiliation{Italian Institute of Technology, 16163 Genova, Italy}

\date{\today}

\begin{abstract}
Supersolid is a mysterious and puzzling state of matter whose possible  existence has stirred a vigorous debate among physicists for over 60 years. Its elusive nature stems from the coexistence of two seemingly contradicting properties, long-range order and superﬂuidity. We report computational evidence of a supersolid phase of deuterium under high pressure ($p >$ 800 GPa) and low temperature (T $<$ 1.0 K). In our simulations, that are based on bosonic path integral molecular dynamics, we observe a highly concerted exchange of atoms while the system preserves its crystalline order. The exchange processes are favoured by the soft core interactions between deuterium atoms that form a densely packed metallic solid. At the zero temperature limit, Bose-Einstein condensation is observed as the permutation probability of $N$ deuterium atoms approaches $1/N$ with a finite superfluid fraction. Our study provides concrete evidence for the existence of a supersolid phase in high-pressure deuterium and could provide insights on the future investigation of supersolid phases in real materials.
\end{abstract}

\maketitle

Reports of an anomalous superfluid phase in solid $^4$He\cite{Kim2004} have spurred renewed interest in the study of this unusual state of matter, often referred to as a supersolid, in which long-range translational order and superfluidity are believed to coexist\cite{THOULESS96, Lifshitz69, Chester1970,Kim2004,day07,Hunt2009,Kreibich2008,Boninsegni12,Yukalov2020,Mezzacapo06,Cinti2010,Tanzi2019,Li2017,Leonard2017}. The very concept of a supersolid is puzzling since in a solid the nuclear density is localised around the equilibrium positions, while in a superfluid the nuclei wavefunctions are delocalised due to exchange \cite{Lifshitz69,Chester1970,Sindzingre1991,Ceperley1995,Mezzacapo06,Cinti2010,Tanzi2019,Li2017,Leonard2017}.  

Theoretical investigations\cite{Penrose1958,THOULESS96, Lifshitz69, Chester1970} have preceded the first experimental reports of a $^4$He supersolid \cite{Kim2004}. Some of them argued that a supersolid could not exist\cite{Penrose1958} while others suggested that defects could favour its formation\cite{THOULESS96, Lifshitz69, Chester1970}. However, the experimental claim of \cite{Kim2004} has been challenged\cite{day07,Hunt2009,Kreibich2008}, and it was pointed out that the defect formation energy in solid $^4$He is too large to be invoked as a pathway to supersolidity\cite{Ceperley2004}. Nevertheless, the search for a supersolid phase has not been abandoned and is still of great interest. Some encouragement in this direction comes from theoretical studies which indicate that a supersolid phase can be stabilized by suitable interparticle interactions\cite{Li2017,Tanzi2019,Cinti2010,Kora2019}, the dimensionality of the system\cite{Mezzacapo06,Cinti2010,Mezzacapo2011} or optical coupling\cite{Leonard2017}.

\begin{figure*}[]
\centering
\subfigure{
\includegraphics[width=16cm]{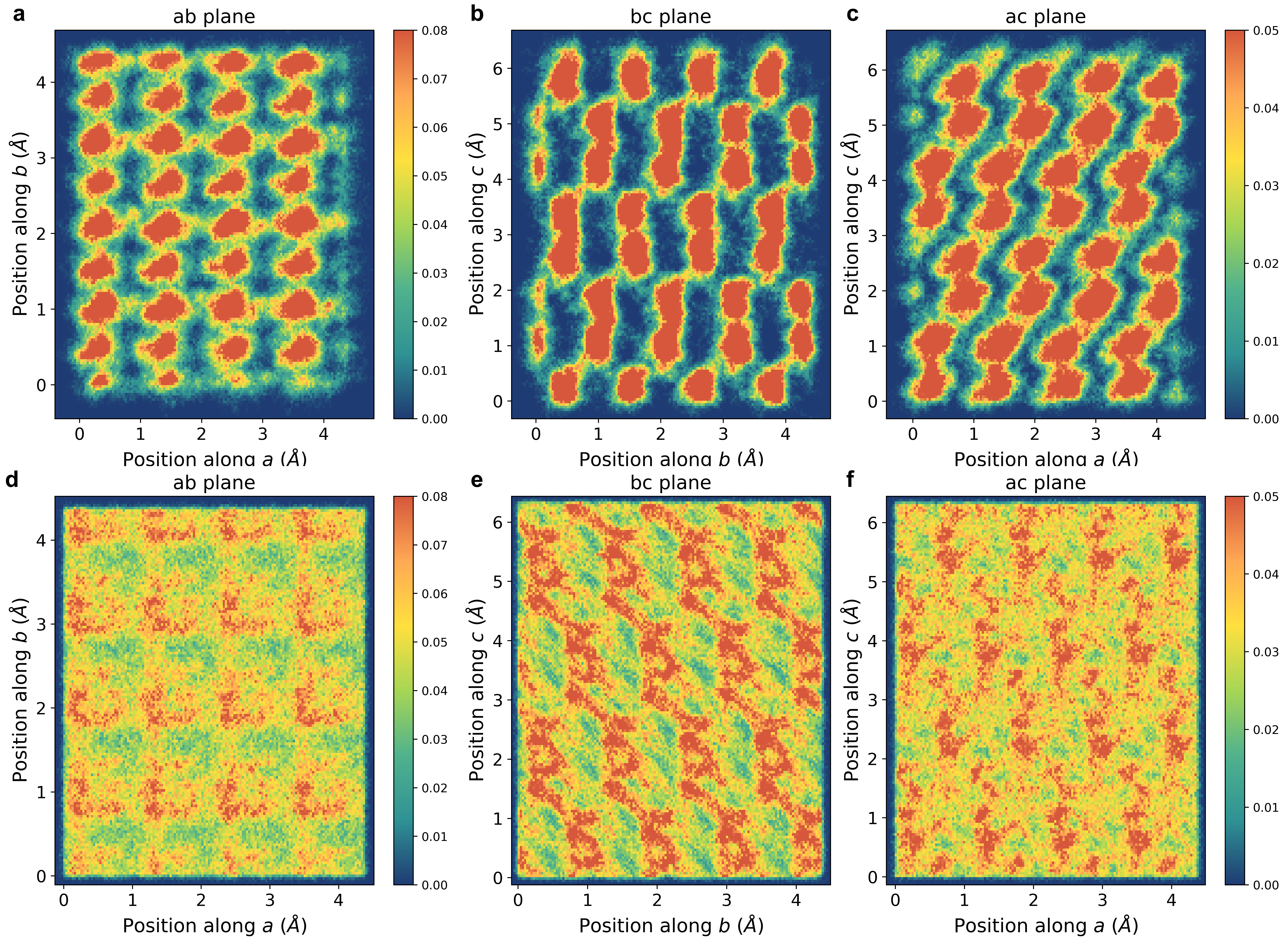}}
\caption{\textbf{Atomic density of high-pressure deuterium solid with exchange.} Two-dimensional (2D) cross sections of the atomic density $n(r)$ of high-pressure deuterium in path integral molecular dynamics (PIMD) (a-c) and bosonic path integral molecular dynamics (PIMD-B) (\textbf{d-f}) simulations at $p=800$ GPa and T = 0.5 K. While ring beads of PIMD simulation are always closed form (a-c), the PIMD-B simulations allow deuterium atoms to exchange (d-f).}
\label{fig:density_nqe}
\end{figure*}

\begin{figure*}[]
\centering
\subfigure{
\includegraphics[width=16 cm]{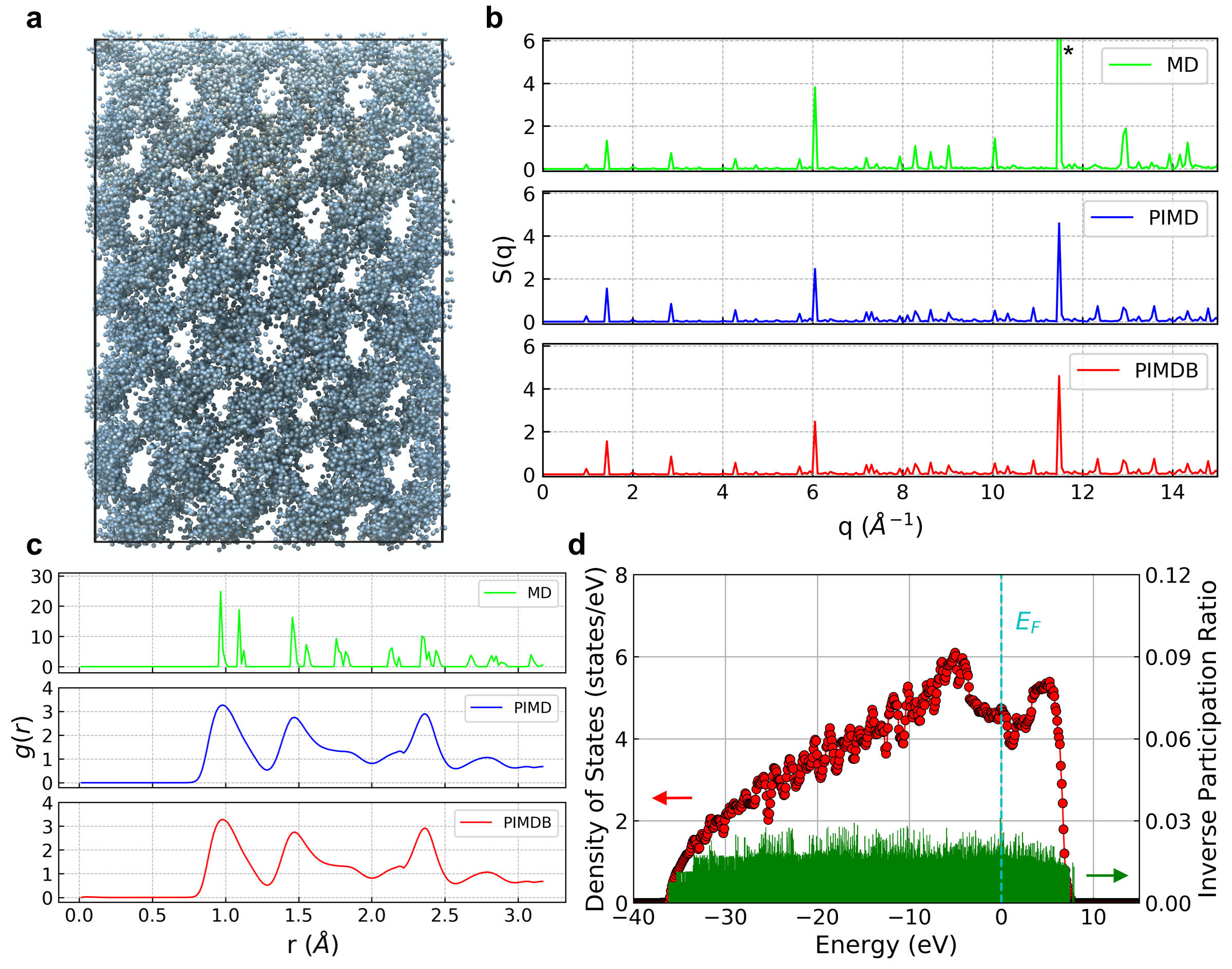}}
\caption{\textbf{Exchange effect on the geometry and electronic properties of high-pressure deuterium.} (a) A snapshot taken from the ($N\times P$) trajectory of PIMD-B simulations at T = 0.5 K and $p = 800$ GPa. Each blue sphere represents the bead of ring polymer (in total $P = 256$ beads and $N=128$). (b) Structure factors $S(q)$ of high-pressure deuterium from the MD (green line), PIMD (blue line) and PIMD-B (red line) simulations. The amplitude of omitted $S(q)$ peak (*) of the MD simulation (green) is 14.9. (c) Radial distribution functions $g(r)$ of high-pressure deuterium from the MD (green line), PIMD (blue line) and PIMD-B (red line) simulations. (d) The density of states (red point) and inverse participation ratio (green bar) of a supersolid phase. The Fermi level ($E_F$) is zero (cyan dashed line).}
\label{fig:geo}
\end{figure*}

\begin{figure}
\includegraphics[width=8cm]{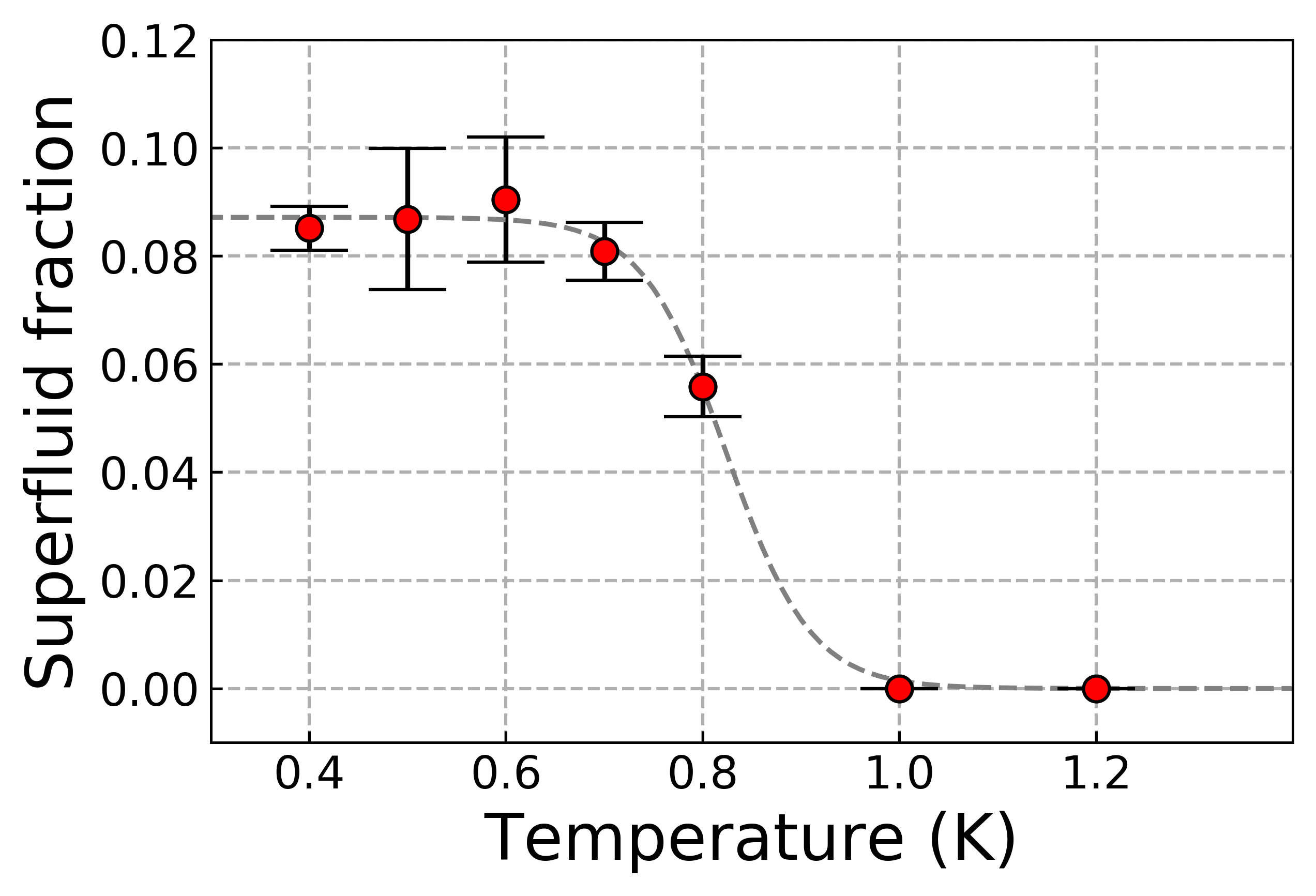}
\caption{\textbf{The superfluid fraction of high-pressure deuterium.} The superfluid fraction (red point with error bar) at $p=800$ GPa as the function of temperature with a guiding line (dotted grey line).}
\label{fig:permutation_prob}
\end{figure}

In this paper, we report numerical evidence that deuterium at low temperature and high pressure can indeed become supersolid.  There are various reasons why we pay attention to high pressure deuterium: Firstly, the light mass of deuterium ($Z=2$) leads to significant nuclear quantum effects (NQEs). Secondly, high level quantum mechanical calculations, such as density functional theory (DFT) and quantum Monte Carlo, predict that deuterium forms a metallic $I4_1/amd$ phase at $p> 500$ GPa\cite{mcmahon11,Azadi14}. Such a compressed environment promotes exchange interactions of deuterium atoms by bringing them closer. Thirdly, it was argued that soft core interatomic potentials aid in favouring a supersolid phase\cite{Cinti2010}. In deuterium, the interactions between the nuclei in the metallic phase have screened Coulomb character which is softer than Lennard-Jones interactions. Lastly, the predicted phase transition pressure of the metallic $I4_1/amd$ phase ($p> 500$ GPa) appears to be within reach of experimental capabilities in the near future\cite{Dalladay-Simpson2016,Ji2019,Dias715,nature20}.

Simulating the quantum behaviour of a supersolid phase of deuterium poses several challenges, such as: 1) the accurate modelling of the interaction potential, 2) the inclusion of NQEs and 3) the introduction of bosonic exchange symmetry. Here we sketch the main points of our approach and refer the interested reader to a more detailed description of our methodology in the Supplemental Material.  Following the approach pioneered by Behler and Parrinello\cite{Behler2007}, the interaction potential is described by a feed forward  neural network potential, that is  trained on a large number of DFT calculations. We chose the vdW-DF2 functional based on the generalized gradient approximation with non-local correlations\cite{Thonhauser2007} (Supplemental Material section I).

NQEs are described by using a discretised version of Feynman’s path integral expression for the quantum partition function that is sampled in molecular dynamics simulations (PIMD)\cite{Parrinello84} by exploiting its well-known isomorphism with a system of classical ring polymers\cite{Chandler1981}. Exchange symmetry is dealt with using the bosonic version of path integral molecular dynamics (PIMD-B) of Hirshberg et al.\cite{Hirshberg2019, Hirshberg2020}. This is done by evaluating the PIMD potential for $N$ bosons recursively,

\begin{equation}
\label{eq:pimdb}
e^{-\beta V^{(N)}_B}=\frac{1}{N}\sum^N_{k=1}e^{-\beta(E^{(k)}_N+V^{(N-k)}_B)},
\end{equation}
where $\beta$ is the inverse temperature, $V^{(N-k)}_B$ is the PIMD potential for $N-k$ bosons and $E^{(k)}_N$ is the spring energy of a ring polymer constructed by connecting all of the beads of $k$ particles sequentially\cite{Hirshberg2019}.
The method provides the correct bosonic thermal expectation values while avoiding the need to enumerate all $N!$ permutations of identical particles. This reduces the computational scaling of bosonic PIMD simulations from factorial to cubic, allowing large bosonic systems to be simulated using PIMD\cite{Hirshberg2019}. We have explicitly checked that this method~\cite{Hirshberg2020} gives results in full agreement with those obtained using the PIMC method pioneered by Ceperley\cite{Ceperley1995}. Our evaluation of superfluid fractions of liquid $^4$He\cite{Ceperley1995} \ctext[RGB]{255,251,204}{and $hcp$ solid $^4$He}\cite{Ceperley2004} \ctext[RGB]{255,251,204}{concurred with the previous PIMC results (Supplemental Fig. 12 and 13).} \ctext[RGB]{255,251,204}{For deuterium, we note that the current implementation only considers the spatial permutation of a spin-polarized system (Supplemental section III). Thus, our estimation is relevant to a spin-polarized system and might lead to a slight overestimation of the superfluid transition temperature.}

To perform simulations at constant pressure, we implemented the NPT PIMD algorithm and adapted it to use the correct pressure estimator for bosons (Supplemental Material section II). Although we have studied the system at different thermodynamics conditions, here we report the results obtained at $p=800$ GPa in a range of low temperatures from $T=0.1$ K to $T=1.2$ K in the main text. Additional thermodynamic conditions are found in the Supplemental Material. We find that converged results can be obtained if we discretise the Feynman path using $P = 256$ beads (Supplemental Fig. 7).

In order to bring out the role of NQEs and exchange symmetry, we performed simulations of solid deuterium using three diﬀerent methods, treating deuterium as 1) a classical particle (MD), 2) a distinguishable quantum particle (PIMD), and 3) an indistinguishable boson (PIMD-B) (Fig. \ref{fig:density_nqe} and \ref{fig:geo}). The average density $n(r)$ is greatly affected by exchange processes (Fig. \ref{fig:density_nqe}). 

Even for distinguishable deuterium the NQEs make the atomic density distribution of neighbouring atoms overlap (Fig. \ref{fig:density_nqe}a-c). This overlap suggests the possible role of exchange processes. Indeed, as the exchange of deuterium atoms is allowed \textit{via} PIMD-B simulation, it is difficult to spot the precise equilibrium positions of deuterium $I4_1/amd$ phase due to active exchange (Fig. \ref{fig:density_nqe}d-f). This implicates that the connected ring polymers of deuterium atoms emerge at low temperatures (Fig. \ref{fig:density_nqe}d-f). At first sight (Fig. \ref{fig:geo}a), it would appear that the $n(r)$ would correspond to that of a glassy system, however our analysis shows that the $I4_1/amd$ symmetry is hidden but not lost (Fig. \ref{fig:geo}b,c). To show it, we evaluated the structure factor, 

\begin{equation}
\label{eq:sq}
S(q) =\frac{1}{PN}\sum^{P}_\tau\sum^{N}_{j,k}e^{-i\mathbf{q}(\mathbf{R}^{(\tau)}_j-\mathbf{R}^{(\tau)}_k)},
\end{equation}
where $P, N, \mathbf{R}^{(\tau)}_j$ are the number of beads, the number of particles and position of atom $j$ at $\tau$ imaginary time, respectively. Bragg peaks can be clearly seen with and without exchange at the same positions in reciprocal space (Fig. \ref{fig:geo}b and Supplemental Fig. 8). Thus, the result indicates that this peculiar exchange of deuterium does not break the solid long-range order. Also, the pair correlation of solid phase is preserved under exchange interactions as evidenced by the radial distribution function $g(r)$ of MD, PIMD and PIMD-B simulations (Fig. \ref{fig:geo}c). Even in the active exchange regime, the system still remains metallic as the solid phase. This can be understood given that this anomalous deuterium phase preserves the solid long-range order. Thus, the density of states (Fig. \ref{fig:geo}d) is similar to that of solid (Supplemental Fig. 2b). The presence of disorder in a supersolid phase might introduce the localisation of electronic states\cite{Anderson58}. However, our analysis based on inverse participation ratio (IPR) shows that the electronic states of supersolid phase are delocalised (Fig. \ref{fig:geo}d and Supplemental section III).

The fact that one can reconcile long range order and a very active exchange regime remains puzzling also in the Feynman isomorphism. In order to get insight into how this is possible, we look at the beads’ spatial arrangement as it evolves during the simulation where all permutations contribute to the forces on atoms at each time step\cite{Hirshberg2019}. This can be measured by a structure factor of the beads system considered as a set of independent particles $S_{rel}(q) =\frac{1}{PN}\sum^{P}_{\tau,\tau^{\prime}}\sum^{N}_{j,k}e^{-iq(\mathbf{R}^{(\tau)}_j-\mathbf{R}^{(\tau^{\prime})}_k)}$. While the beads distribution changes dynamically from one time step to another, the overall long-range order of $(P \times N)$ configuration is still preserved (Supplemental Fig. 9). This points to a highly coherent exchange mechanism. 

An elegant way of measuring whether a system is superfluid is to compute its winding number\cite{Pollock1987}. This quantity reflects the number of paths that, due to exchange, are so long that they wrap around the periodic boundary conditions\cite{Pollock1987}. In our approach, in which all permutation are sampled at every time step, standard methods to evaluate it cannot be applied. Therefore, we have developed an approximate but highly accurate approach to measure the winding number in PIMD-B simulations (Supplemental Material section III and Supplemental Fig. 11). The result obtained is presented in Fig. \ref{fig:permutation_prob}. It shows that at $T<1.0$ K a superfluid condensate is formed. The analysis of probability of observing longer rings also confirms this picture (Supplemental Fig. 12). Our calculation shows that for high pressure deuterium a defect-free  pathway to supersolidty is possible. 


Experiments on such thermodynamic conditions will be feasible in near future given the rapid advancement of diamond anvil cell techniques at cryogenic temperature\cite{Dias715,Dalladay-Simpson2016,Ji2019}, and verifying this prediction in experiments will be a fascinating challenge to undertake.



\begin{acknowledgments}
We are grateful to L. Bonati, M. Yang, V. Rizzi, D. Frenkel,  V. Kapil, C. Schran and K. Trachenko for helpful discussions. This research was supported by the European Union (Grant No. ERC-2014-ADG-670227/VARMET) and the NCCR MARVEL, funded by the Swiss National Science Foundation. Computational resources were provided by the Euler cluster at ETH Zürich and the Swiss National Supercomputing Centre (CSCS) under project ID s1052. C.W.M. acknowledges the support from Korea Institute of Science and Technology Information (KISTI) for the Nurion cluster (KSC-2019-CRE-0139 and KSC-2019-CRE-0248). Part of this work was performed under the gracious hospitality of ETH Zürich and Universit\`a della Svizzera italiana, Lugano.

All the implementations of the isotropic and full-cell NPT simulations of PIMD-B are freely available in the \texttt{LAMMPS Github} repository. All the necessary input files of this computational study are also available in the author's \href{https://github.com/changwmyung/H_PIMD}{\texttt{Github}} repository. 
\end{acknowledgments}

\bibliography{ms}
\end{document}


\title[\today]{Supplemental Material: Prediction of a supersolid phase in high-pressure deuterium}

\author{Chang Woo Myung}
\affiliation{Yusuf Hamied Department of Chemistry, University of Cambridge, Lensﬁeld Road, Cambridge, CB2 1EW, United Kingdom}

\author{Barak Hirshberg}
\email{hirshb@tauex.tau.ac.il}
\affiliation{School of Chemistry, Tel Aviv University, Tel Aviv 6997801, Israel}

\author{Michele Parrinello}
\email{michele.parrinello@iit.it}
\affiliation{Italian Institute of Technology, 16163 Genova, Italy}

\date{\today}

\maketitle

\tableofcontents

\pagebreak

\section{Density functional calculations and machine learning potential}

\subsection{Convergence of density functional theory calculations}

\floatsetup[figure]{style=plain,subcapbesideposition=top}
\begin{figure}[h]
\includegraphics[width=0.45\linewidth, height=5cm]{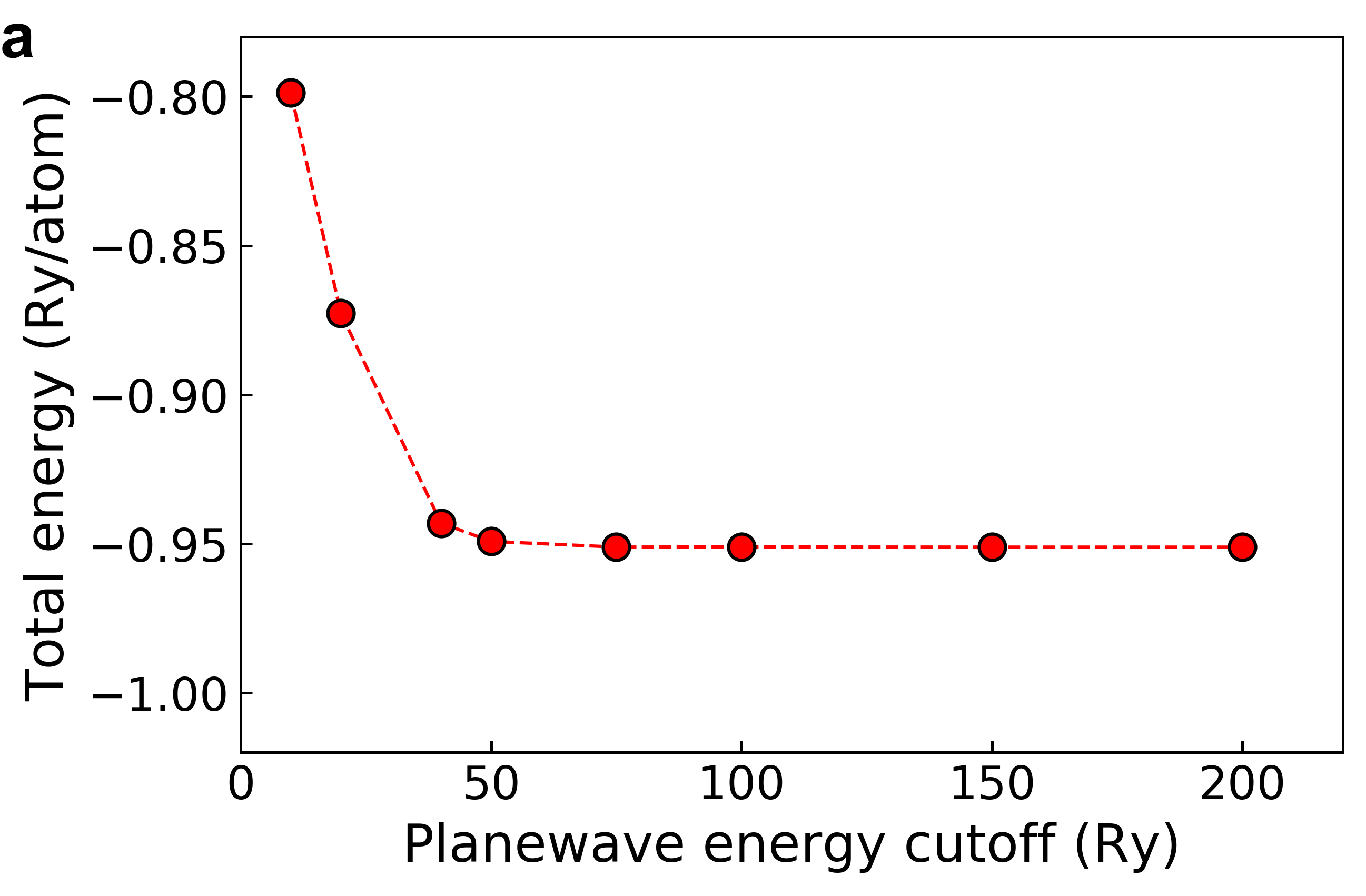} 
\label{fig:ecut}
\includegraphics[width=0.45\linewidth, height=5cm]{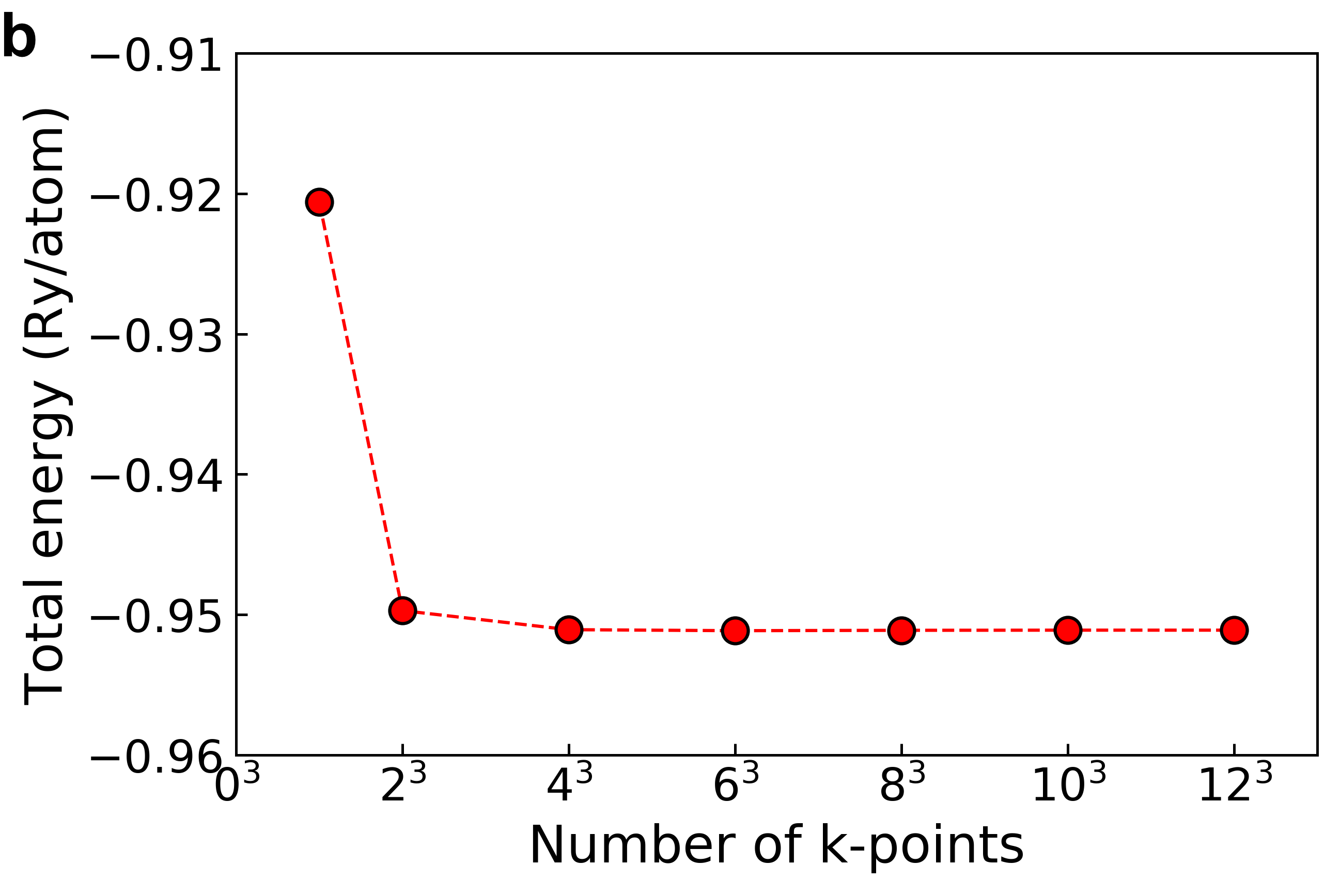}
\label{fig:kp}
\caption{\textbf{The convergence test of DFT total energy.} The convergence of the total energy per atoms with respect to (a) the planewave energy cutoff and (b) the number of k-points.}
\label{fig:conv}
\end{figure}

Since the interatomic distance between deuterium atoms in high-pressure solid phases is small ($<1.0 \AA$), the reliability of density functional theory (DFT) calculation is affected by the pseudisation radius ($r_c$) of PAW pseudopotential. In the pressure range of $p = 800-1200 \ \text{GPa}$, the Wigner-Sitz radius of system ($r_s$) is $\sim 0.5$ \AA ($\sim 0.94 \ a.u.$). Previous DFT study showed that the $r_c$ of $0.5 a.u.$ is required to ensure the convergence of total energy of high-pressure hydrogen solid\cite{mcmahon11}. Therefore, we employ a PAW pseudopotential of $r_c =0.5 \ a.u.$ following the previous work. 

As the long-range van der Waals (vdW) interactions play important roles in high-pressure hydrogen/deuterium systems\cite{li2013,cui20}, we calculated the energy, forces and stress tensors using the non-local vdW functional, vdW-DF2 functional\cite{Thonhauser2007, Berland2015, Sabatini2012}, implemented in \texttt{Quantum Espresso} package (v6.6)\cite{PGiannozzi2017}. A previous study of $I4_1/amd$ phase showed that the long-range dispersion contribution to the enthalpy is sensitive to the size of a supercell. They observed that the long-range contribution starts to converge from a supercell of 72 atoms\cite{cui20}. Therefore, we used a supercell of 128 atoms to eliminate any finite size effects related to the long-range dispersion interactions.

Because the high-pressure deuterium system is highly compressed with a small unit-cell, rigorous tests are needed to ensure the total energy convergence depending on planwave energy cutoff (Supplemental Fig.  \ref{fig:conv}a) and the number of k-points(Supplemental Fig.  \ref{fig:conv}b). We choose the planewave energy cutoff of $100 \ \text{Ry}$ that shows $0.8 \ \text{meV/atom}$ error compared to the fully converged case of $200 \ \text{Ry}$. We use the k-point mesh of ($10\times10\times8$) that shows $0.01 \ \text{meV}$ error compared to the fully converged case of k-mesh ($12\times12\times12$) in Supplemental Fig. \ref{fig:conv}b. The corresponding $k$-grid spacing is 2$\pi \times$ 0.0136 \AA$^{-1}$. This is finer than the k-mesh spacing of previous studies of metallic hydrogen (2$\pi \times$ 0.04 \AA$^{-1}$\cite{Pickard2007} or 2$\pi \times$ 0.05 \AA$^{-1}$ \cite{Cudazzo08}) to ensure the convergence of Fermi surface. Therefore, we train the machine learning (ML) potentials for molecular dynamics simulations based on the DFT calculations with above parameters.

\subsection{Electronic and vibrational structures of deuterium $I4_1/amd$ phase}

\begin{figure}[h]
\includegraphics[width=0.45\linewidth, height=5cm]{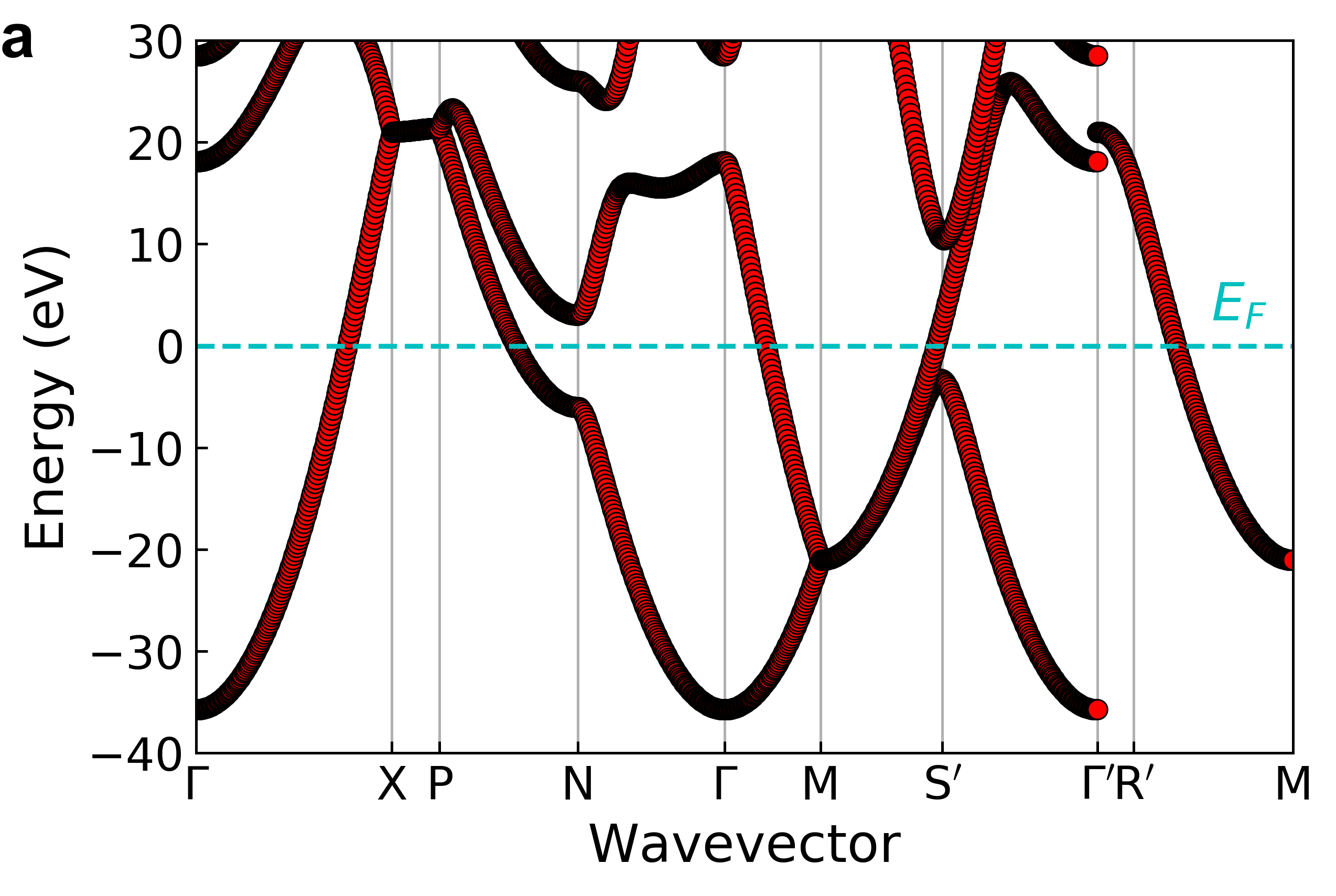} 
\label{fig:band}
\includegraphics[width=0.45\linewidth, height=5cm]{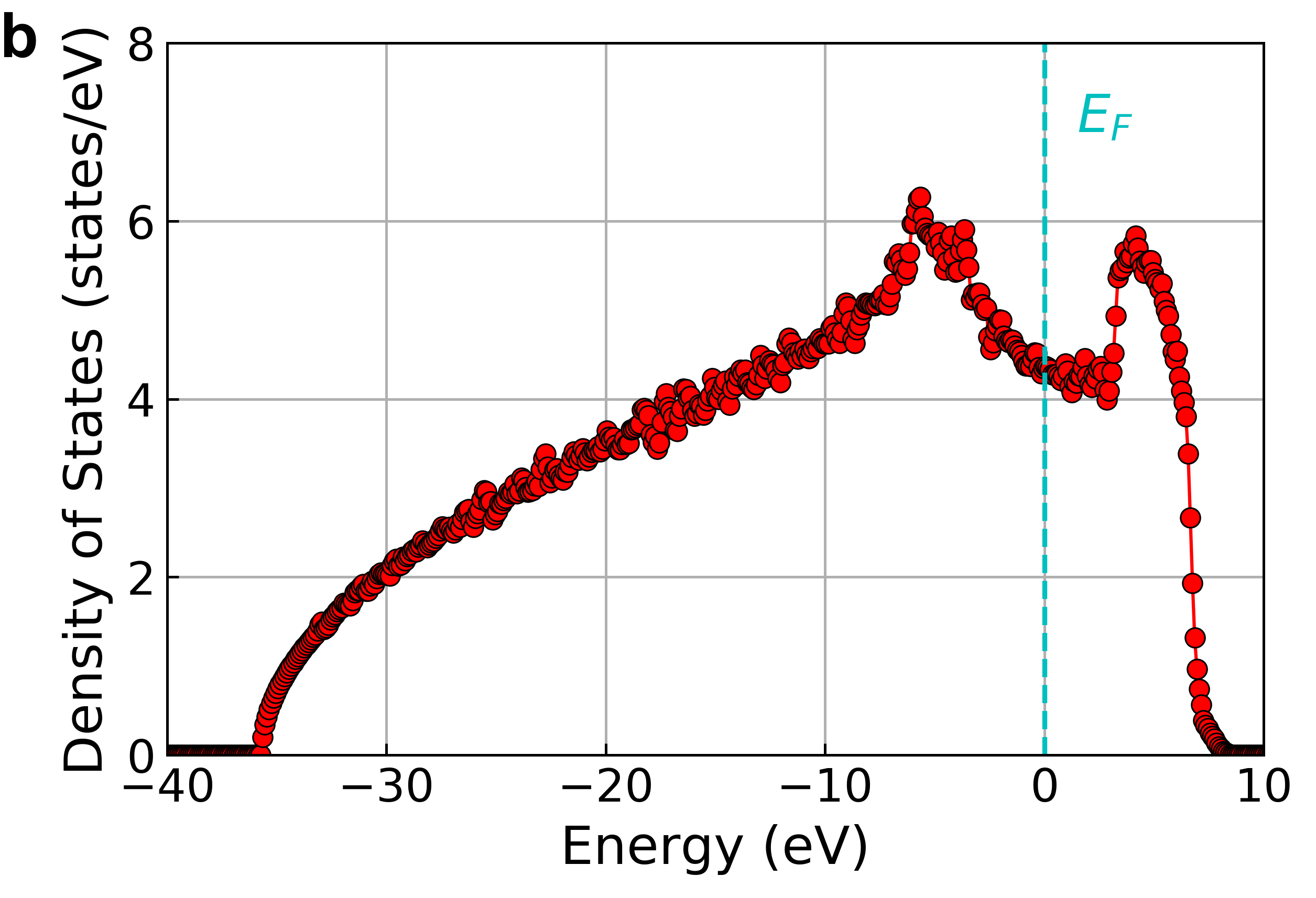}
\label{fig:dos}
\caption{\textbf{The electronic structure of deuterium $I4_1/amd$ solid phase.} (a) The band structure of deuterium $I4_1/amd$ solid phase of the primitive cell, where $S'=S|S_0$, $\Gamma'=\Gamma|X$ and $R'=R|G$ and high-symmetry points $\Gamma$=(0.0,0.0,0.0), X=(0.0,0.0,0.5), P=(0.25,0.25,0.25), N=(0.0,0.5,0.0), M=(0.5, 0.5,-0.5), S=(0.28, 0.72, -0.28), S$_0$=(-0.28, 0.28, 0.28), R=(-0.06, 0.06, 0.5), G=(0.5, 0.5, -0.06) of the BZ. (b) The density of states of hydrogen $I4_1/amd$ solid phase of ($4\times4\times2$) supercell. The fermi energy ($E_F$) is indicated as cyan dashed line.}
\label{fig:elec}
\end{figure}

Our DFT calculations of band structure (Supplemental Fig.  \ref{fig:elec}) and density of states (Supplemental Fig.  \ref{fig:elec}) at the vdW-DF2 level are consistent with the previous works in which the $I4_1/amd$ phase is metallic . 

\floatsetup[figure]{style=plain,subcapbesideposition=top}

\begin{figure}[h]
\includegraphics[height=6cm]{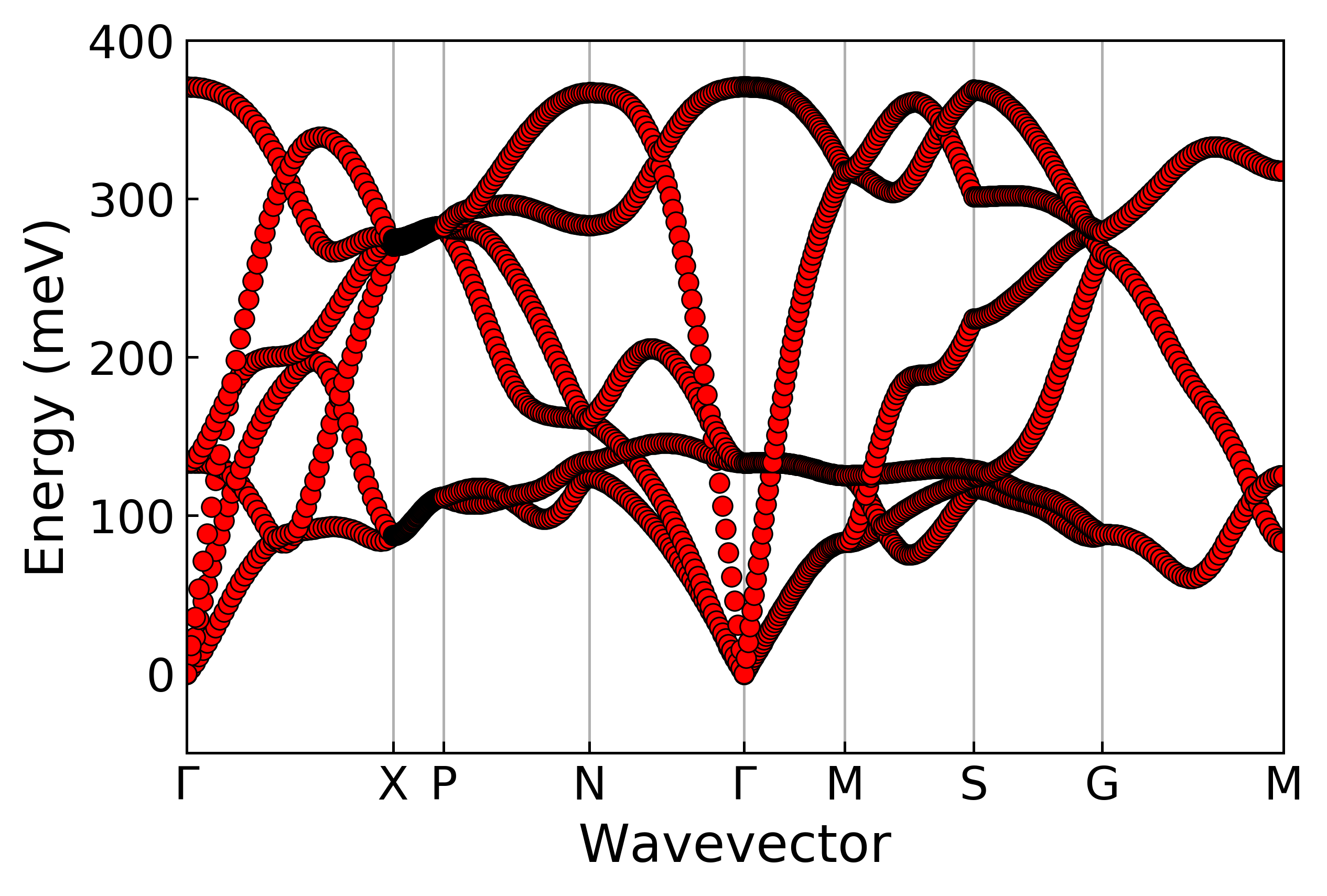} 
\caption{\textbf{The phonon dispersion of deuterium $I4_1/amd$ phase.} The phonon band structure of hydrogen $I4_1/amd$ solid phase of $(4\times4\times4)$ supercell of the primitive cell, where high-symmetry points are $\Gamma$=(0.0,0.0,0.0), X=(0.0,0.0,0.5), P=(0.25,0.25,0.25), N=(0.0,0.5,0.0), M=(0.5, 0.5,-0.5), S=(0.28, 0.72, -0.28), R=(-0.06, 0.06, 0.5), G=(0.5, 0.5, -0.06) of the BZ.}
\label{fig:ph}
\end{figure}

It is also important to ensure that our DFT calculation describes the vibrational property of $I4_1/amd$ phase accurately without any unstable modes. This is crucial for building accurate ML potentials to perform molecular dynamics simulations. The phonon dispersion of $I4_1/amd$ phase has no imaginary phonon branch across the Brillouin zone (BZ) (Supplemental Fig.  \ref{fig:ph}). 

\floatsetup[figure]{style=plain,subcapbesideposition=top}

\subsection{Machine learning potential}

\floatsetup[figure]{style=plain,subcapbesideposition=top}
\begin{figure}[h]
\includegraphics[height=6cm]{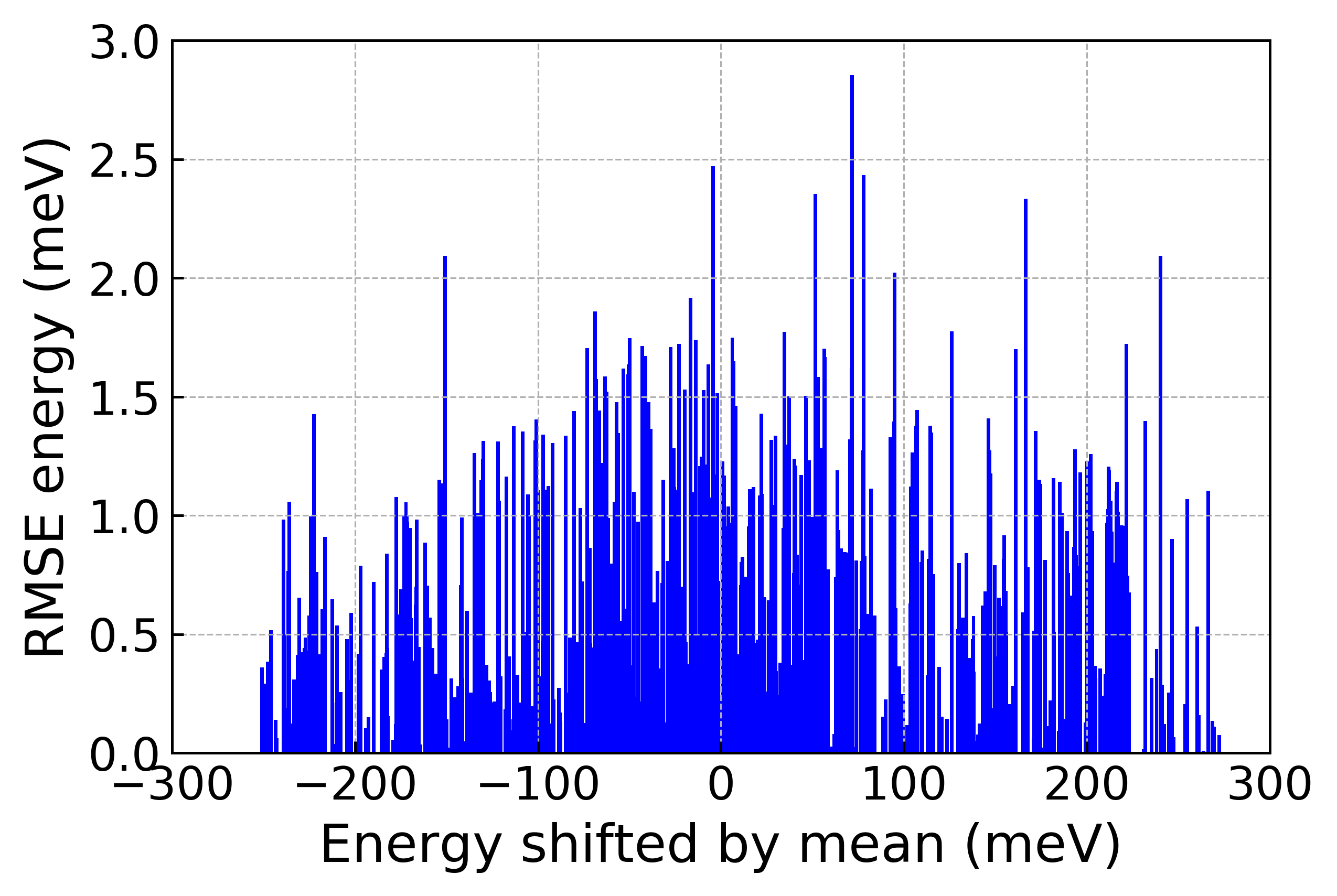} 
\caption{\textbf{The total energy root mean squared error (RMSE) of ML potential for the testing set.} The total energy RMSE per atom (meV/atom) is plotted by bars as a function of total energy (meV/atom) shifted by its average, 12.687 eV/atom.}
\label{fig:errore}
\end{figure}

\floatsetup[figure]{style=plain,subcapbesideposition=top}
\begin{figure}[h]
\includegraphics[height=6cm]{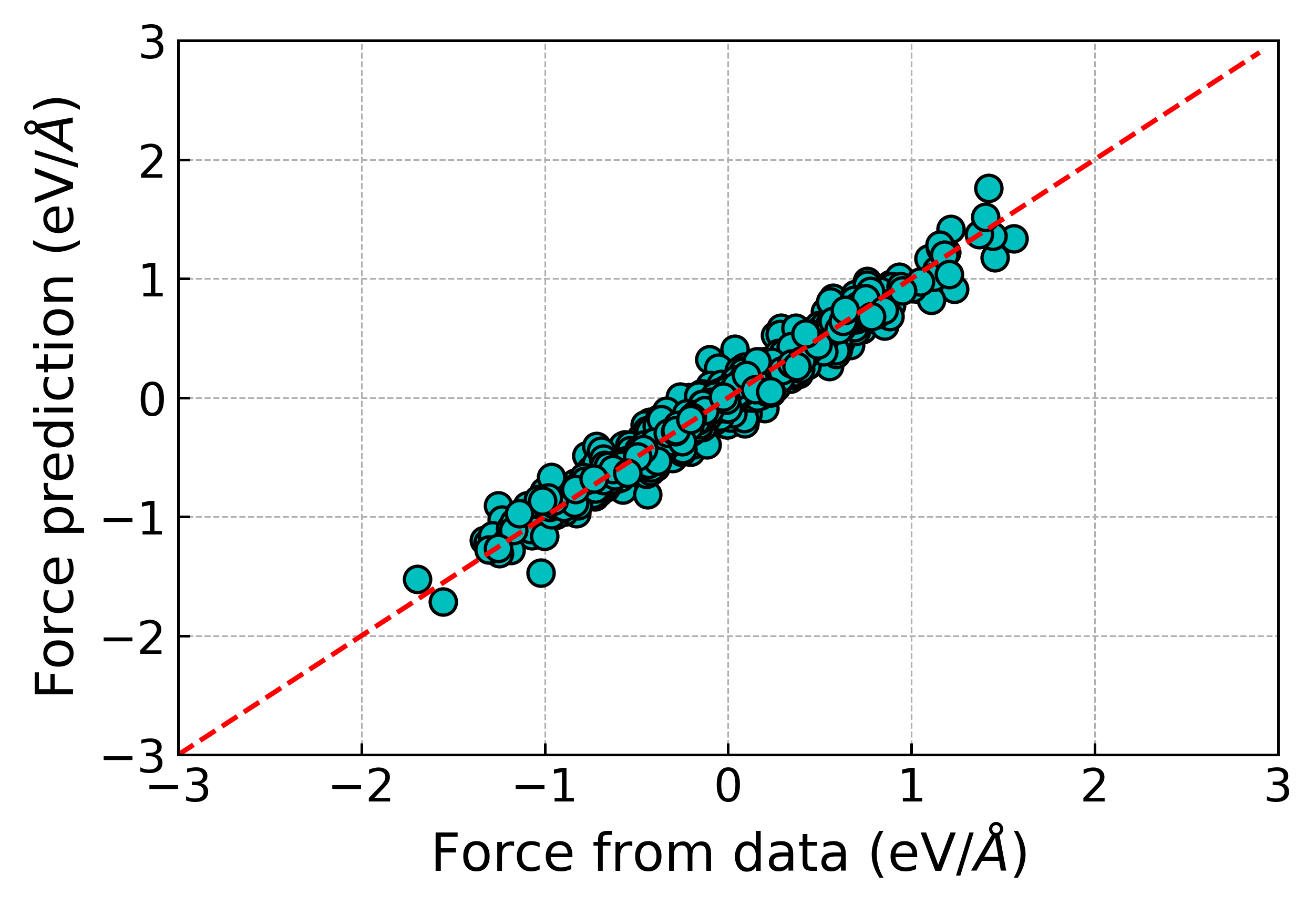} 
\caption{\textbf{The comparison of forces between DFT and ML potential for the testing set.} The $x,y,z$ components of forces $(f_x,f_y,f_z)$ of DFT and ML potential for the testing set are plotted.}
\label{fig:errorf}
\end{figure}

\floatsetup[figure]{style=plain,subcapbesideposition=top}
\begin{figure}[h]
\includegraphics[height=6cm]{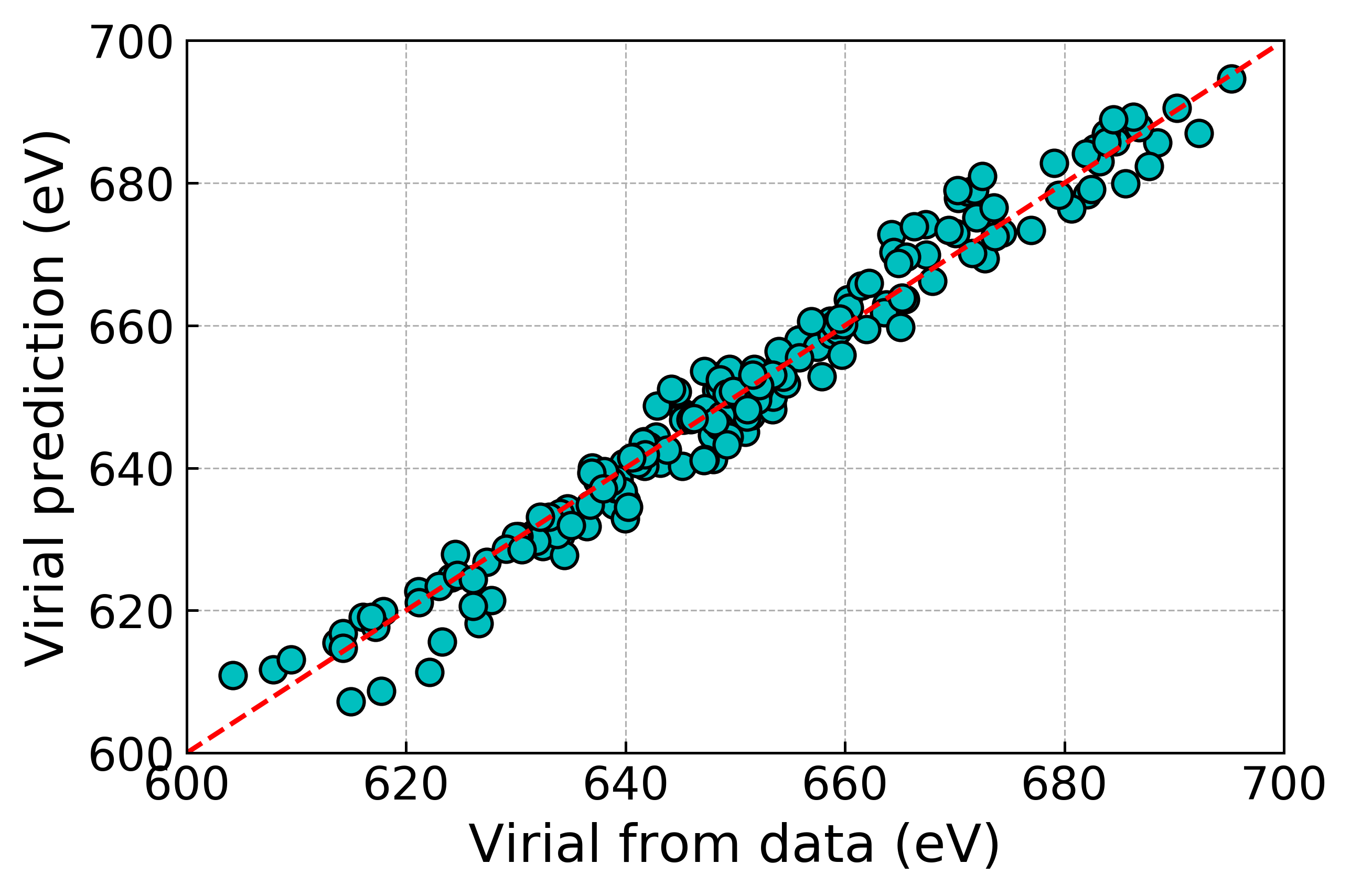} 
\caption{\textbf{The comparison of stress tensor virials between DFT and ML potential for the testing set.} The comparison of stress tensor virial $(v_{xx}, v_{yy}, v_{zz}, v_{yz}, v_{xz}, v_{xy})$ of DFT and ML potential for the testing set are plotted for the case of $p = 800$ GPa.}
\label{fig:errorv}
\end{figure}

\captionsetup{format=naturestyle,labelformat=naturestyle-table,justification=just,singlelinecheck=false,labelsep=naturestyle}

\begin{table}[]
\begin{tabular}{cccc}
\hline
\hline
Pressure (GPa) & Energy ($meV/$atom) & Force ($meV/$\AA) & Virial ($eV$) \\ \hline
800            & 0.7                 & 60.0              & 2.48        \\
1000           & 0.8                 & 86.6              & 1.93        \\
1200           & 1.0                 & 91.1              & 2.05             \\ \hline \hline
\end{tabular}
\caption{\label{tab:mlp-error}The test errors of neural network potential for potential energy ($meV/atom$), force ($meV/\AA$) and pressure virial (eV) at $p=800, 1000, 1200$ GPa.}
\end{table}

\captionsetup{format=naturestyle,labelformat=naturestyle,justification=just,singlelinecheck=false,labelsep=naturestyle}

We sampled the configurations of solid and liquid phases of high-pressure deuterium by using Born-Oppenheimer DFT (vdW-DF2 functional\cite{Thonhauser2007, Berland2015, Sabatini2012}) NPT MD simulations at the temperature range of $0.5 \ \text{K}<T<600 \ \text{K}$ and the pressure range of $800-1200 \ \text{GPa}$ with the $k$-mesh of $(6\times6\times4)$. By randomly choosing $\sim 50k$ configurations from the DFT MD trajectories, we construct a first ML potential. We used this potential to run preliminary PIMD-B simulations. We selected from these simulations $\sim 50k$ configurations (including long exchange ones) for which we recalculated energy,  forces  and  stress  virials  using  a denser $k$-mesh of $(10\times10\times8)$. We trained the final ML potential using these data.

We trained ML neural network potential for each pressure (800, 1000 and 1200 GPa) at the whole temperature range using the \texttt{DeepMD-kit} package\cite{Wang2018} with smooth edition (SE) descriptor\cite{Zhang2018}. The SE descriptor is constructed to represent  the atomic environment  by $(32\times64\times128)$ neural network with 16 axis neurons\cite{Zhang2018}. The energy, forces and tensor virials are  predicted by the fitting neural network of 4 hidden layers ($512\times256\times128\times64\times32$) over $1-10$ million iterations. The test errors at $p=800, 1000, 1200$ GPa are found in Table 1. For instance, the test errors of 800 GPa case for the total energy (Supplemental Fig.  \ref{fig:errore}), force (Supplemental Fig.  \ref{fig:errorf}) and stress tensor virial (Supplemental Fig.  \ref{fig:errorv}) are 0.7 meV/atom 60.0 meV/$\AA$ and 2.48 eV, respectively. 

\subsection{Defect}
From the early stage, defect has been considered as the most plausible pathway in forming a supersolid phase\cite{THOULESS96, Lifshitz69, Chester1970,Pederiva97,Ceperley04,day07}. In $^4$He solid, however, PIMC simulations showed that the formations of vacancy and interstitial are thermodynamically unfavourable\cite{Pederiva97,Ceperley04}. Since a defect-free supersolid phase was not observed in a PIMC simulation, the origin of a $^4$He supersolid phase is still under debate\cite{day07}. In light of previous studies, the role of defect as a pathway to supersolid should not be overlooked in high-pressure deuterium. Therefore, we investigate the thermodynamic stability of various defect types in high-pressure deuterium solid.

Because the atomic positions of $I4_1/amd$ phase are all symmetrically equivalent, only a single type of mono vacancy exists. On the other hands, there exist the two types of interstitials, $D_{3h}$ and $D_{4h}$.

The single point DFT calculation shows that the formation energies of the above defects are too high to be formed compared to the melting temperature of high-pressure deuterium $\sim 120$ K. The mono deuterium vacancy defect has the formation energy of $E[\text{V}^0_{\text{D}}] = 647 \ \text{meV}$. The $D_{4h}$ interstitial formation energy is $E[\text{D}_i] = 280 \ \text{meV}$. And the $D_{3h}$ interstitial is highly unstable that any local minimum configuration is not found. Finally, the formation energy of vacancy-interstitial pair is  $E[\text{V}^0_{\text{D}}+\text{D}_i] = 286 \ \text{meV}$. Therefore, we conclude that the mono and pair defects of the $I4_1/amd$ phase does not exist at $T\sim 1$ K. 

\section{\label{sec:level1} NPT implementation of bosonic path integral molecular dynamics}

Our implementation of NPT PIMD-B simulation follows the NPT PIMD algorithm of Martyna et al. where we adopt the Nose-Hoover chain thermostat/barostat\cite{Martyna99,martyna96}. The major revision of PIMD-B simulation on the algorithm is the inclusion of bosonic exchange to the pressure estimator. 

\subsection{\label{sec:level3} Primitive pressure estimator}

The primitive pressure estimator $p^{(prim)}_{\alpha\beta}$ of PIMD is given by
\begin{eqnarray}
\nonumber
p^{(prim)}_{\alpha\beta} = &\frac{NPk_BT}{det(\overrightarrow{h})} \delta_{\alpha\beta} + \frac{1}{det(\mathbf{h})} \sum_i^N \Big(\Phi_{spring}(\textbf{r}^{(\tau)}_i) \\
&+\frac{1}{P}\sum^P_\tau (\textbf{f}^{(\tau)}_i)_\alpha (\textbf{r}^{(\tau)}_i)_\beta \Big)- \sum_\mu^d \sum^P_\tau h_{\beta\mu} \frac{\phi(\mathbf{h}, \mathbf{r}^{(\tau)})}{\partial h_{\alpha\mu}},
\end{eqnarray}
where $\alpha(\beta), N,P,T,\mathbf{r}^{(\tau)}_i,\mathbf{f}^{(\tau)}_i$ and $\mathbf{h}$ are Cartesian axis, the number of particles, the number of beads, temperature, the position of the particle $i$ at imaginary time $\tau$, the force on the particle $i$ at imaginary time $\tau$ and the cell matrix. The interactions between beads at different imaginary times $\tau$ are given as the spring term $\Phi_{spring}(\mathbf{r}^{(\tau)}_i)$. In PIMD NPT simulation without bosonic exchange, $\Phi_{spring}(\mathbf{r}^{(\tau)}_i)= -m_i\omega^2_P\sum^P_\tau (r^{(\tau)}_i-r^{(\tau)}_{i+1})_\alpha (r^{(\tau)}_i-r^{(\tau)}_{i+1})_\beta$. If the nuclei follow Bose statistics, the pressure virial should be modified since the forces on atoms need to include exchange effects, which we discuss in the following subsection \ref{sec:level3BosonP}. 

\subsection{\label{sec:level3BosonP} Derivation of pressure estimator of indistinguishable Bosonic NPT simulation}

The quantum partition function $Q_P$ of the 3D system of $N$ particles with $P$ beads,
reciprocal temperature $1/k_BT$ and cell shape $\textbf{h}$ is given by

\begin{eqnarray}
\nonumber
Q_P(N,\beta,h) &= \Big[ \prod^N_{i} \Big(\frac{m_i P k_B T}{2\pi\hbar^2}\Big)^{3P/2} \Big] \int dr^{(1)}_i ... dr^{(P)}_i \\
& e^{-\beta[\sum^{N,P}_{i,\tau}\Phi_{spring}(r^{(\tau)}_i) + \frac{1}{P}\sum^P_\tau \phi(r^{(\tau)}_i,V)]}
\end{eqnarray}

The spring term under bosonic exchange is\cite{Hirshberg2019}
\begin{eqnarray}
\Phi_{spring} = V^{(N)}_B = -\frac{1}{\beta} ln \Big[ \frac{1}{N} \sum^N_{k=1} e^{-\beta(E^{(k)}_N+V^{(N-k)}_B)} \Big]
\end{eqnarray}

As $Q_P$ depends the cell matrix $\mathbf{h}$, the pressure tensor \(p_{\alpha\beta}\) under full-fledged cell fluctuations is 
\begin{eqnarray}
p_{\alpha\beta} = \frac{1}{\beta det[h]} \sum^{x,y,z}_{\gamma} h_{\beta\gamma} \Big( \frac{\partial ln Q_M}{\partial h_{\alpha\gamma}} \Big)_{N,T}
\end{eqnarray}

We adapt the scaled variable $\mathbf{s}_i$ of atom $i$ in the cell $\mathbf{h}$ as $\mathbf{r}_i = \mathbf{h} \cdot \mathbf{s}_i$. In the summation form, the $\alpha$ component position of atom $i$ is $r_{i,\alpha} = \sum_\beta h_{\alpha\beta}s_{i,\beta}$. Since we are interested in the spring term, which now considers the
bosonic symmetry, we focus on the spring contribution to the pressure
tensor.

The spring term contribution to the pressure tensor becomes (in scaled coordinates),
\begin{eqnarray}
p_{\alpha\beta, spring} = \sum^{x,y,z}_{\gamma} \sum^{N,P}_{i,\tau} \frac{\partial \Phi_{spring}}{\partial(h \cdot s^{(\tau)}_i)}_{\alpha} h_{\beta\gamma} s^{(\tau)}_{i,\gamma}.
\end{eqnarray} 
where $\mathbf{h \cdot s^{(\tau)}_i} =
r^{(\tau)}_i$. If we consider distinguishable particles, converting it into the Cartesian coordinates results in
\begin{eqnarray}
p_{\alpha\beta, spring} = \sum^{N,P}_{i,\tau} - m_i\omega_P (r^{(\tau+1)}_i-r^{(\tau)}_i)_\alpha (r^{(\tau+1)}_i-r^{(\tau)}_i)_\beta.
\end{eqnarray}

With the bosonic symmetry, the spring contribution becomes
\begin{eqnarray}
\label{eq:p_force}
p_{\alpha\beta, spring}^{(N)} &= \sum^{N,P}_{i,\tau} \frac{\partial V^{(N)}_B}{\partial r^{(\tau)}_{i,\alpha}} r^{(\tau)}_{i,\beta}\\
\label{eq:p_iter}
&= \frac{\sum^N_{k=1} \Big[ \sum^{N,P}_{i,\tau} \Big( \frac{\partial E^{(k)}_N}{\partial r^{(\tau)}_{i,\alpha}} r^{(\tau)}_{i,\beta} + \frac{\partial V^{(N-k)}_B}{\partial r^{(\tau)}_{i,\alpha}} r^{(\tau)}_{i,\beta} \Big) \Big] e^{-\beta(E^{(k)}_N+V^{(N-k)}_B)}}{\sum^N_{k=1} e^{-\beta(E^{(k)}_N+V^{(N-k)}_B)}}
\end{eqnarray}
And we note that
\begin{eqnarray}
\sum^{N,P}_{i,\tau} \frac{\partial E^{(k)}_N}{\partial r^{(\tau)}_{i,\alpha}}r^{(\tau)}_{i,\beta} = -m_i\omega_P \sum^N_{i=N-k+1}\sum^P_{\tau} (r^{(\tau+1)}_i-r^{(\tau)}_i)_\alpha (r^{(\tau+1)}_i-r^{(\tau)}_i)_\beta
\end{eqnarray}
It is implied that $r^{(M+1)}_N = r^1_{N-k+1}$ or otherwise $r^{M+1}_i = r^{(1)}_{i+1}$. The spring contribution to the pressure tensor is calculated through the iterative equation (\ref{eq:p_iter}). 

However, since we calculate the forces of spring term $\frac{\partial V^{(N)}_B}{\partial r^{(\tau)}_{i,\alpha}}$ already, the spring contribution to the pressure virial is obtained by equation (\ref{eq:p_force}). Although we have implemented both isotropic and full-cell Parrinello-Rahman NPT PIMD-B, we only present the isotropic NPT. All the features are implemented in a development version of \texttt{LAMMPS} and be found in the author's \href{https://github.com/changwmyung/H_PIMD}{\texttt{Github}} repository.

\subsection{\label{sec:level3}Equations of motion for NPT PIMD-B simulation}

We follow the NPT equations of motion of Martyna et al. where each Cartesian degree of freedom of the system, $dN$, couples to the Nose-Hoover chain (NHC)\cite{Martyna99,martyna96}, where $d$ is the dimension of system. Each Nose-Hoover chain couples to the each degree of freedom $dNPN_{nhc}$, where $N_{nhc}$ is the number of NHC. The default number of NHC for thermostat/barostat are $N_{nhc}=3$ for the whole PIMD-B calculations in this work. The only difference of PIMD-B NPT simulation compared to PIMD is that now the bosonic pressure estimator is used to measure the pressure of system. And although NPT PIMD simulation usually uses center of mass (centroid) pressure estimator and the corresponding equations of motion, the definition of centriod of ring polymers becomes elusive and ill-defined with the bosonic symmetry. Therefore, NPT PIMD-B simulation calculates primitive pressure estimator of equation (\ref{eq:p_force}) and the corresponding equations of motion without introducing the centroid.

\begin{eqnarray}
\dot{u}_i^{(\tau)} = &\frac{\mathbf{p}_i^{(\tau)}}{m_i^{(\tau)}} + \frac{\mathbf{p}_\epsilon}{W} u_i^{(\tau)},\\
\dot{\mathbf{p}}_i^{(\tau)} = &\mathbf{g}_i^{(\tau)} + \frac{1}{P} \mathbf{f}_i^{(\tau)} - (1+\frac{1}{NP})\frac{p_\epsilon}{W} \mathbf{p}_i^{(\tau)} - \frac{\mathbf{b}_{\xi_{i1, i}}^{(\tau)}}{Q_{i,1}^{(\tau)}},\\
\dot{V} = &\frac{dVp_\epsilon}{W},\\
\dot{p}_\epsilon = & dV(p_{int}-p_{ext})+\frac{1}{NP}\sum_I^N\sum_\tau^P\frac{(\mathbf{p}_i^{(\tau)})^2}{m_i^{(\tau)}}-\frac{p_{\eta_1}}{Q_1^{\epsilon}}p_\epsilon,\\
\dot{\xi}_{i,k}^{(\tau)}= &\frac{\mathbf{p}_{\xi_{i,k}}^{(\tau)}}{Q_{i,k}^{(\tau)}},\\
\dot{\eta}_k^{(\tau)}=&\frac{p_{\eta_k}}{Q_k^{(\epsilon)}},\\
\dot{\mathbf{p}}_{\xi_{i,1}}^{(\tau)}=&\big[ \frac{\mathbf{b}_{i,i}^{(\tau)}}{m_i^{(\tau)}} - k_B T \hat{n} \big] - \frac{\mathbf{b}_{\xi_{i,1},\xi_{i,2}}^{(\tau)}}{Q_{i,2}^{(\tau)}},\\
\dot{\mathbf{p}}_{\xi_{i,2}}^{(\tau)}=& \big[ \frac{\mathbf{b}_{\xi_{i,1},\xi_{i,1}}^{(\tau)}}{m_{i,1}^{(\tau)}} - k_B T \hat{n} \big],\\
\dot{p}_{\eta_1}=& \big[ \frac{p^2_\epsilon}{W} -k_B T] - \frac{p_{\eta_2}}{Q_2^{(\epsilon)}} p_{\eta_1},\\
\dot{p}_{\eta_2}=&\big[ \frac{(p_{\eta_{1}})^2}{Q^{(\epsilon)}_{1}} -k_BT].
\end{eqnarray}
where $\mathbf{u}^{(\tau)}_i$ is the normal mode transformation of the position of particle $i$ in the imaginary time $\tau$, $p_\epsilon$ is the momentum conjugate to $\epsilon=lnV$, $\eta_k$ is the $k$th element of the volume thermostat whose momentum conjugate is $p_{\eta k}$. $\xi^{(\tau)}_{i,k}$ is the $k$th element of the thermostat chain of the particle $i$ in the imaginary time $\tau$. And $\mathbf{p}^{(\tau)}_{\xi_{i,k}}$ is its conjugate momentum. The barostat mass parameters are $W=d(NM+1)k_BT/\omega^2_b$ and $Q_k^{(\epsilon)}=\frac{k_BT}{\omega^2_b}$, where $\omega_b$ is the damping frequency. The vectors $\mathbf{b}^{(\tau)}_{\xi,\xi'}$ and $\mathbf{n}$ are [($p^{(\tau)}_{\xi,x}p^{(\tau)}_{\xi',x}$), ($p^{(\tau)}_{\xi,y}p^{(\tau)}_{\xi',y}$), ($p^{(\tau)}_{\xi,z}p^{(\tau)}_{\xi',z}$)] and [1,1,1], respectively.

In PIMD-B NPT equations of motion, we account for the bosonic symmetry by calculating the spring forces of beads $\mathbf{g}^{(\tau)}_i=-\nabla_{u_i^{(\tau)}} \Phi_{spring}(\mathbf{u}_i^{(\tau)})$ and bosonic pressure estimator $p_{int}$ that includes bosonic exchange effects. 

\subsection{Convergence of Bosonic NPT path integral molecular dynamics}
The Bosonic PIMD NPT simulations at various thermodynamics conditions ($0.6 \ \text{TPa} < \text{P} < 1.2 \ \text{TPa}$ and $0.1 \ \text{K} < \text{T} < 500 \ \text{TPa}$) are performed with the time step of $\Delta t = 0.5 \ fs$. Throughout the whole simulations, we used ensemble sampling frequency (damping parameter) for thermostat $\omega = 100\times \Delta t = 50 \ fs^{-1}$ and for barostat $\omega_b = 2000 \times \Delta t = 1.0 \ ps^{-1}$.

\floatsetup[figure]{style=plain,subcapbesideposition=top}
\begin{figure}[h]
\includegraphics[width=0.45\linewidth, height=5cm]{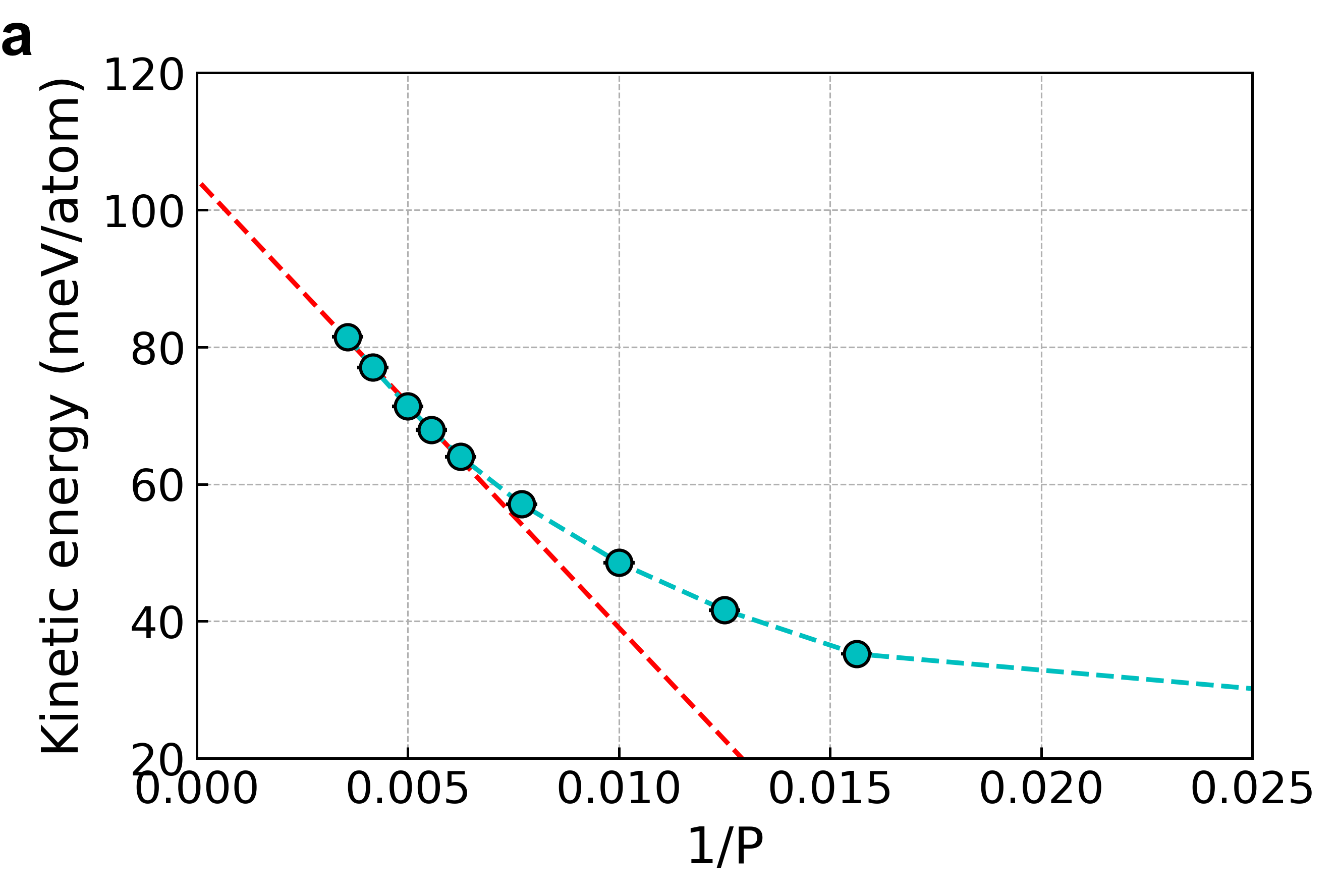} 
\label{fig:ekin}
\includegraphics[width=0.45\linewidth, height=5cm]{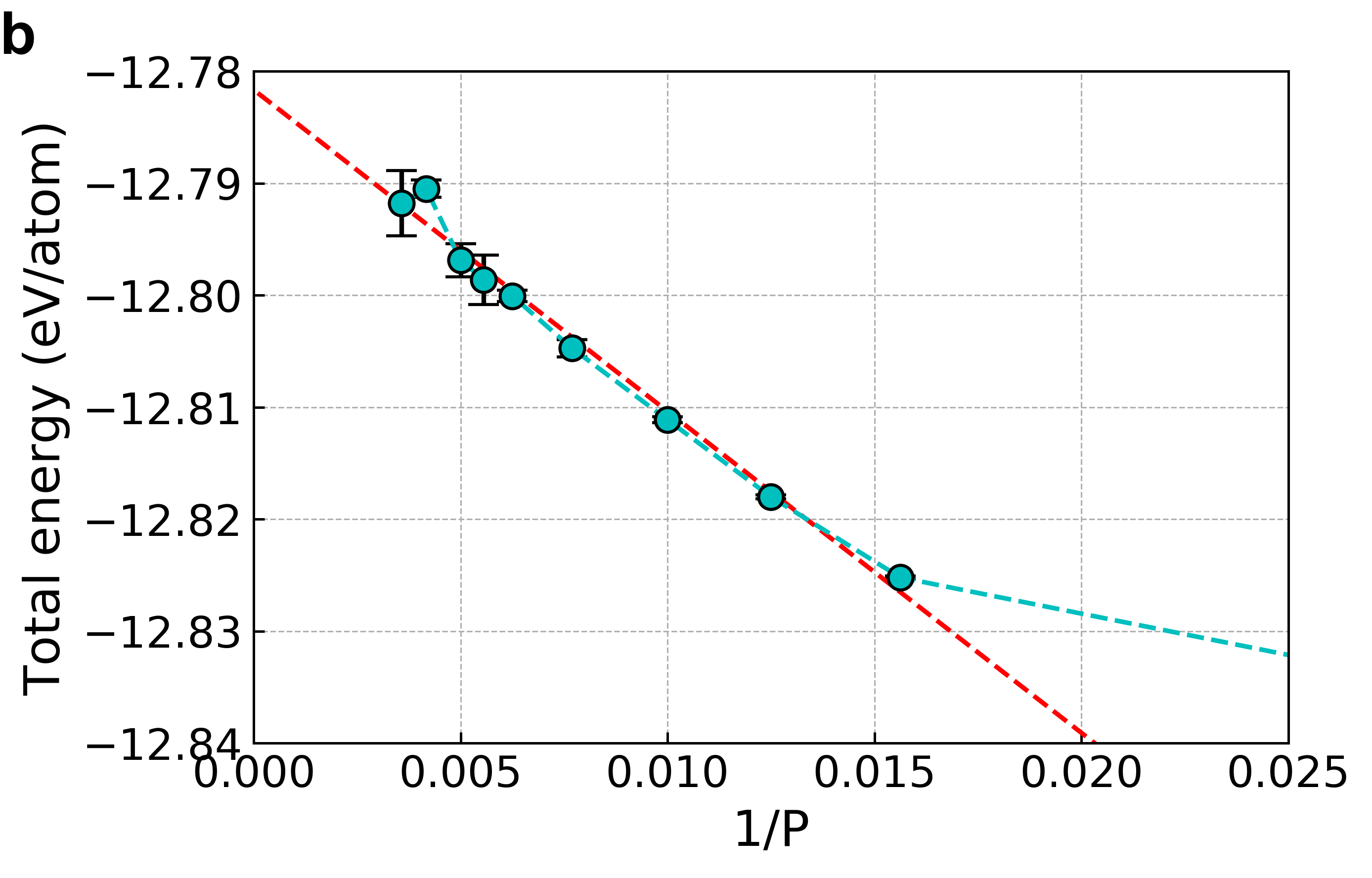}
\label{fig:etot}
\caption{\textbf{The convergence of NPT PIMD-B simulation with respect to the inverse of the number of beads $1/P$.} The convergence of (a) the kinetic energy and (b) the total energy of NPT PIMD-B simulation as the function of $1/P$ at $T=5$ K and $P=800$ GPa. The fitting line (red dashed line) indicates the extrapolated kinetic and total energy at the limit of $P \to \infty$.}
\label{fig:conv-pimd}
\end{figure}
The potential energy, kinetic energy and total energy of Bosonic PIMD NPT at $T = 5$ K and $p=800$ GPa are measured with respect to the number of beads $P =$ 4, 8, 16, 64, 80, 100, 130, 160, 180, 200, 240, 280 (Supplemental Fig.  \ref{fig:conv-pimd}). At $P \sim 250$, 80 \% of kinetic energy (Supplemental Fig. \ref{fig:conv-pimd}a) and 0.1 \% of the total energy (Supplemental Fig. \ref{fig:conv-pimd}b) are converged. 

\floatsetup[figure]{style=plain,subcapbesideposition=top}
\begin{figure}[h]
\includegraphics[height=6cm]{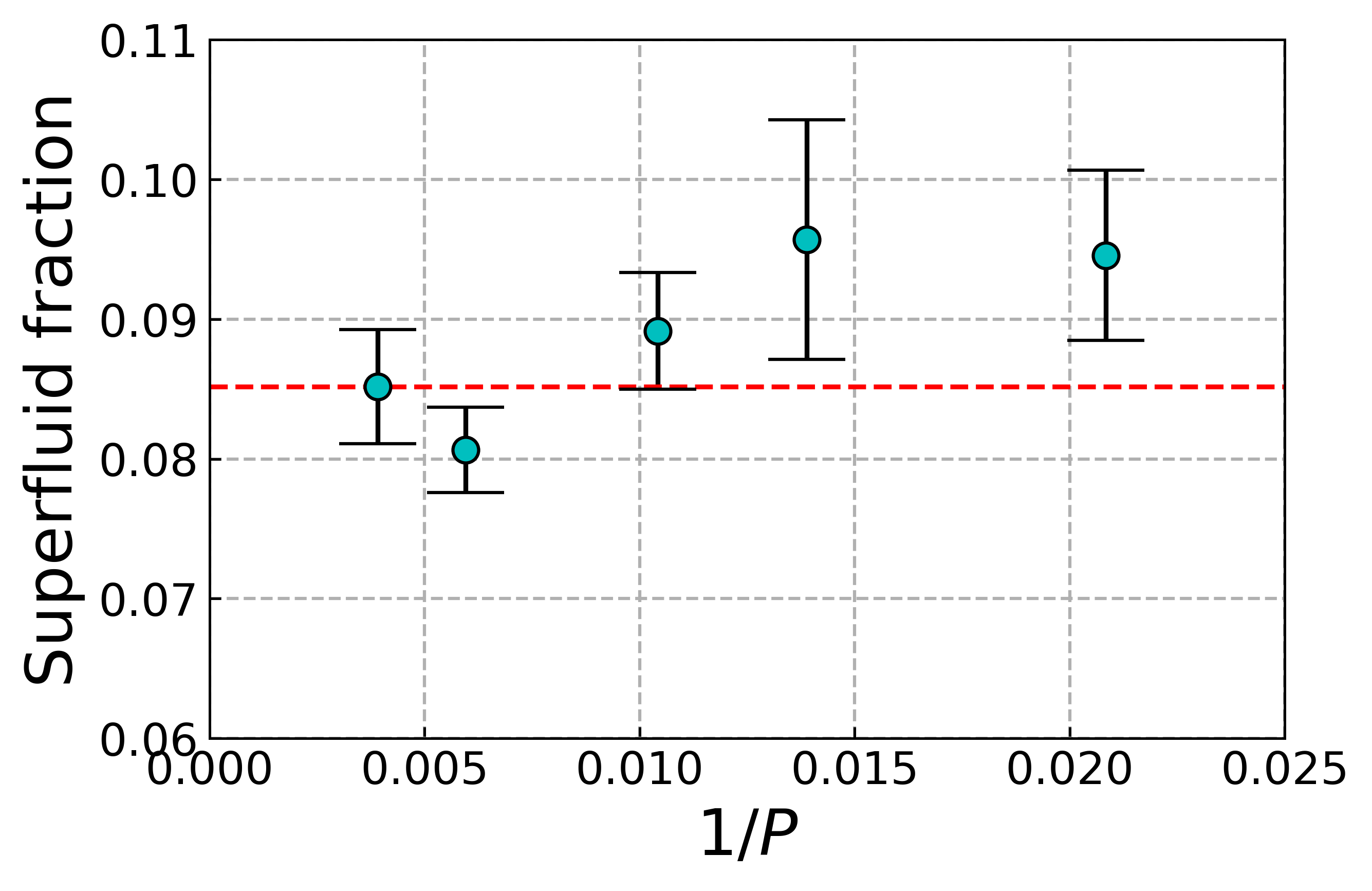} 
\caption{\textbf{The convergence of superfluid fraction with respect to the inverse of the number of beads $\mathbf{1/P}$ at $\mathbf{p=800}$ GPa and $\mathbf{T=0.4}$ K.} The convergence of superfluid fraction with respect to the inverse of number of beads (1/P) with error bars (cyan). The red dashed line indicates the average of superfluid fraction density at $P=256$.}
\label{fig:rho_s_conv}
\end{figure}

In addition, we test the convergence of superfluid fraction (Supplemental Fig. \ref{fig:rho_s_conv}) which will be discussed in detail in Section \ref{winding}. We note that the superfluid density measured at $p=800$ GPa and $T=0.4$ K converges well against $P=256$. Therefore, we used $P=256$ beads for PIMD-B simulation in this work. 

\section{Bosonic path integral molecular dynamics}

In principle, one should account for the permutation of spin coordinates in addition to the spatial coordinates for deuterium atoms which are spin 1 Bosons. Including the spin variables is in principle possible but extremely difficult\cite{Lyubartsev93}, and often ignored such that the system is considered as a spin-polarized system. Ceperley pointed out that factoring the wavefunction to a spin-polarized one would complicate the analysis of rotational symmetry but is not known to cause problems in extended many-body systems\cite{Ceperley91}. \ctext[RGB]{255,251,204}{Following this argument, we neglect the permutation of spin coordinates and only permutes the positions. However, we note that this assumption might lead to a slight overestimation of the superfluid transition temperature}.



\subsection{Calculation of structural properties}

\begin{figure}[h]
\includegraphics[height=6cm]{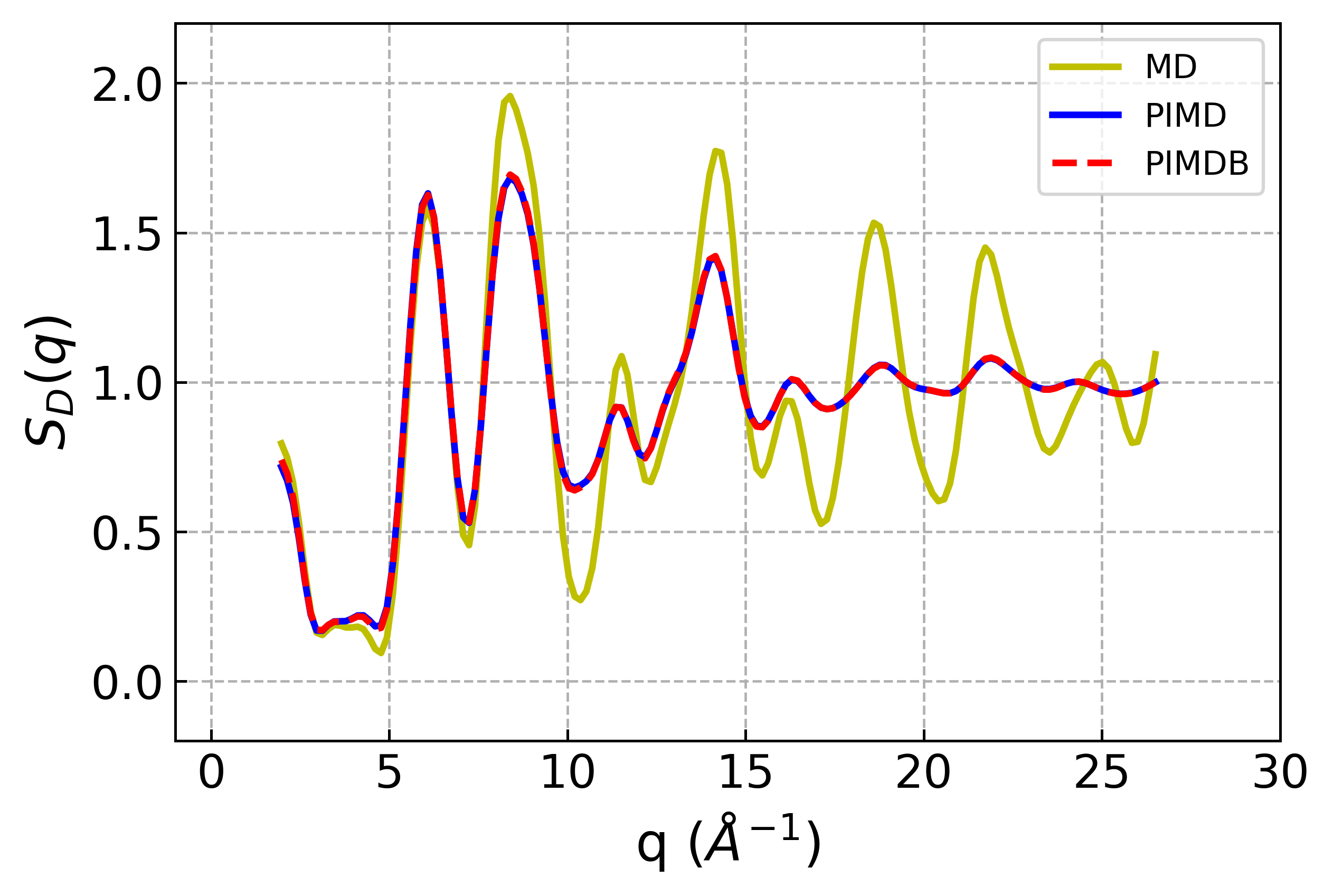} 
\caption{\textbf{Debye structure factors at various of levels of theories.} The Debye structure factor $S_D(q)$ of the system in MD (yellow line), PIMD (blue line) and PIMD-B (red dashed line) simulations.}
\label{fig:Sqdebye}
\end{figure}

The Debye structure factors $S_D(q)$ of MD, PIMD and PIMD-B simulations (Supplemental Fig.  \ref{fig:Sqdebye}) reveal that the peaks of $S_D(q)$ match for all three levels of theories except for the short range ($q > 22 \AA^{-1}$). This result provides that the crystalline long-range order of solid is maintained with dominant exchange of nuclei in PIMD-B simulation. The structure factor was calculated based on the analytic atomic scattering factor\cite{Colliex2004} with parameterized coefficients\cite{Peng96}. 

\floatsetup[figure]{style=plain,subcapbesideposition=top}

\begin{figure*}[]
\centering
\includegraphics[width=16cm]{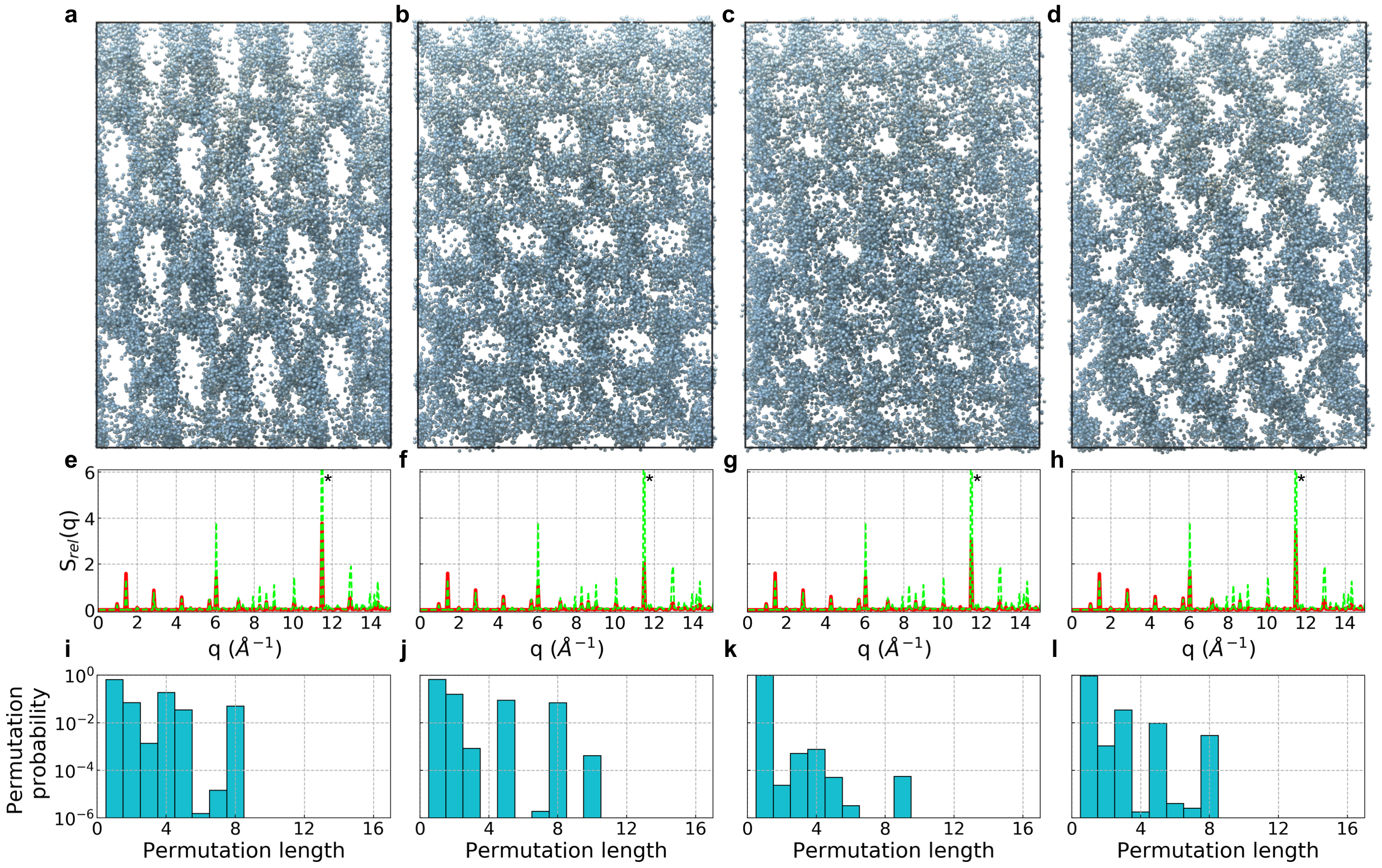}
\caption{\textbf{An ensemble of $\mathbf{P\times N}$ configurations in PIMD-B simulation.} (a-d) Sampled $(P\times N)$ configurations at different time steps in which dynamic exchange occurs. A blue sphere in the frame represents each $(P\times N)$ bead. (e-h) The relative structure factors $S_{rel}(q)$ measured at each time step (red line) referenced to the MD simulation (green dashed line). The amplitude of omitted $S_{rel}(q)$ peak (*) in the MD simulation is 14.9. (i-l) Instantaneous permutation probability of the system at a given time step.}
\label{fig:relSq}
\end{figure*}

In PIMD-B, at each time step, all permutations contribute to the force on each atom\cite{Hirshberg2019}. Therefore, the method can provide insight on how the sampled configurations of $P$ beads of $N$ deuterium atoms evolve in time ($P=256$ and $N=128$) (Supplemental Fig.   \ref{fig:relSq}a-d). Along with the sampled configuration, we show a relative structure factor, $S_{rel}(q) =\frac{1}{PN}\sum^{P}_{\tau,\tau^{\prime}}\sum^{N}_{j,k}e^{-iq(R^{(\tau)}_j-R^{(\tau^{\prime})}_k)}$ (Supplemental Fig.   \ref{fig:relSq}e-h). Although this is not the physical quantum structure factor, this rather describes the properties of the ensemble of $(P \times N)$ configurations sampled at any given time step. In spite of dynamic change in $(P\times N)$ distribution, the overall long-range order of $(P\times N)$ configuration is maintained (Supplemental Fig.   \ref{fig:relSq}e-h). The corresponding instantaneous permutation probability also fluctuate in time while maintaining its overall shape of the average permutation probability (Supplemental Fig.   \ref{fig:relSq}i-l). 




\subsection{Particle indices shuffle in PIMD-B simulation}

\begin{figure}[]
\centering
\includegraphics[width=16cm]{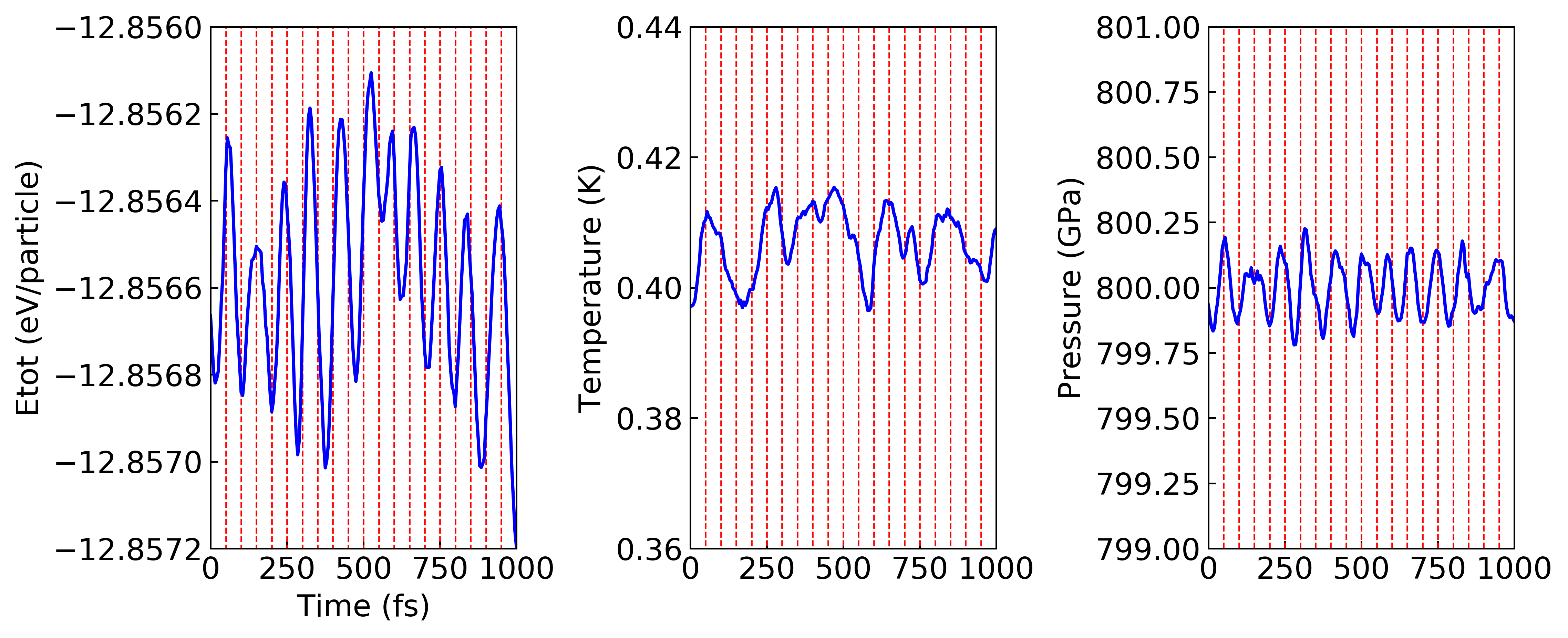}
\caption{\textbf{Indices shuffling effect on thermodynamic quantities} (a) The total energy (eV/particle), (b) temperature and (c) pressure of the system at T = 0.4 K and $p=800$ GPa (blue line). Particle indices shuffling occurs every $N_s=100$ steps (50 $fs$), which is indicated by red vertical dashed lines.}
\label{fig:shuffle}
\end{figure}

Although particle permutation in PIMD-B simulation converges for enough simulation time, we develop a particle indices shuffling scheme for better permutation sampling given limited simulation time. Compared to the default recursive summation approach, the order of particle indices in the summation (equation 1 of main text) is randomly shuffled at every $N_s$ step. In our simulation, we used $N_s=100$ in which thermodynamic quantities are sufficiently equilibrized. Upon the indices shuffling, the thermodynamic observables, such as total energy, temperature and pressure, are not affected ensuring that the system is not pushed out of equilibrium (Supplemental Fig.   \ref{fig:shuffle}).

\subsection{Winding number analysis in PIMD-B simulation} \label{winding}

\IncMargin{1em}
\begin{algorithm}
\SetKwData{Connectivity}{connectivity}\SetKwData{NNlist}{nnlist}\SetKwData{Winding}{W}\SetKwData{This}{this}\SetKwData{Up}{up}\SetKwData{Left}{left}
\SetKwFunction{Union}{Union}\SetKwFunction{CompareDistance}{CompareDistance}\SetKwFunction{FindCompress}{FindCompress}\SetKwFunction{Update}{Update}\SetKwFunction{Connect}{Connect}\SetKwFunction{CalculateWinding}{CalculateWinding}
\SetKwInOut{Input}{input}\SetKwInOut{Output}{output}
\Input{PIMD-B $(P\times N)$ trajectories}
\Output{Average Winding number}
\BlankLine
\emph{Iterate over the sampled time frame}\;
\For{$t\leftarrow 1$ \KwTo $t_{max}$}{
  \emph{Iterate over the number of particle $N$}\;
  \For{$i\leftarrow 1$ \KwTo $N$}{\label{forins}
\Connectivity$\leftarrow$ \CompareDistance{$(r^{(P-1)}_i-r^{(0)}_i), (r^{(P-1)}_i-r^{(0)}_j)$}\;
\If(\tcp*[h]{connect particle $i$ and $j$}){$i$ is not $j$}{\label{lt}
      (\Connectivity,\NNlist)$\leftarrow$ \Update()
      }
\Else(\tcp*[h]{particle $i$ is closed}){(\Connectivity,\NNlist)$\leftarrow$ \Update()}
\If(\tcp*[h]{revert the exchange}){$j$ is in \NNlist}{\label{lt}
      (\Connectivity,\NNlist)$\leftarrow$ \Update()}

}
\Connectivity$\leftarrow$\Connect() \tcp*[h]{close any open rings}\;
\Winding$\leftarrow$\CalculateWinding() \tcp*[h]{calculate Winding number}\;
}
\caption{Winding number calculation in PIMD-B simulation}
\label{algo_winding}
\end{algorithm}\DecMargin{1em}

In a periodic system, the winding number \(\mathbf{W}\) analysis provides facile calculation of superfluid fraction $\rho_s/\rho$ in the path integral method\cite{Pollock1987}. 

\begin{eqnarray}
\mathbf{W}\mathbf{L}= \sum^N_i (\mathbf{r}^{(P-1)}_i-\mathbf{r}^{(0)}_i)
\end{eqnarray} where N is the number of particles, P is the number of beads, $\mathbf{W}$ is the winding number and $\mathbf{L}$ is the unit-cell. However, the PIMD-B simulation calculates the force only, and it is not possible to obtain the exact permutation configuration directly from the trajectories, unlike the PIMC simulation. 

To circumvent this problem, we attempt to identify the permutation configuration by comparing the distances between $(\mathbf{r}^{(P-1)}_i-\mathbf{r}^{(0)}_i)$ and $(\mathbf{r}^{(P-1)}_i-\mathbf{r}^{(0)}_j)$ (equation (\ref{eq:min_fn})),

\begin{eqnarray}
min\Big[|\mathbf{r}^{(P-1)}_i-\mathbf{r}^{(0)}_i|, |\mathbf{r}^{(P-1)}_i-\mathbf{r}^{(0)}_j|\Big].
\label{eq:min_fn}
\end{eqnarray}
If the nearest neighbour of the last bead of a particle $i$, is the first bead of the same particle $i$, it is reasonable to assume that no exchange occurs. On the other hands, if the nearest neighbour of the last bead of a particle $i$ is the first bead of the other particles $j$,  $\mathbf{r}^{(0)}_j$, then we assume that the particle exchange occurs between the particle $i$ and $j$. The estimation is reasonable given that the ring-polymer potential is a harmonic function $ \propto (r^{(\tau+1)}_i-r^{(\tau)}_i)^2$. Algorithm \ref{algo_winding} ensure that the permutation configurations form closed loops. The superfluid fraction $\rho_s/\rho$ is given as
\begin{eqnarray}
\frac{\rho_s}{\rho} = 2\pi \sum_{\alpha=x,y,z} \Big[ \Big( \frac{L_\alpha}{\lambda_D} \Big)^2 \frac{<W_\alpha^2>}{3N} \Big].
\end{eqnarray}

\floatsetup[figure]{style=plain,subcapbesideposition=top}
\begin{figure}[h]
\includegraphics[height=6cm]{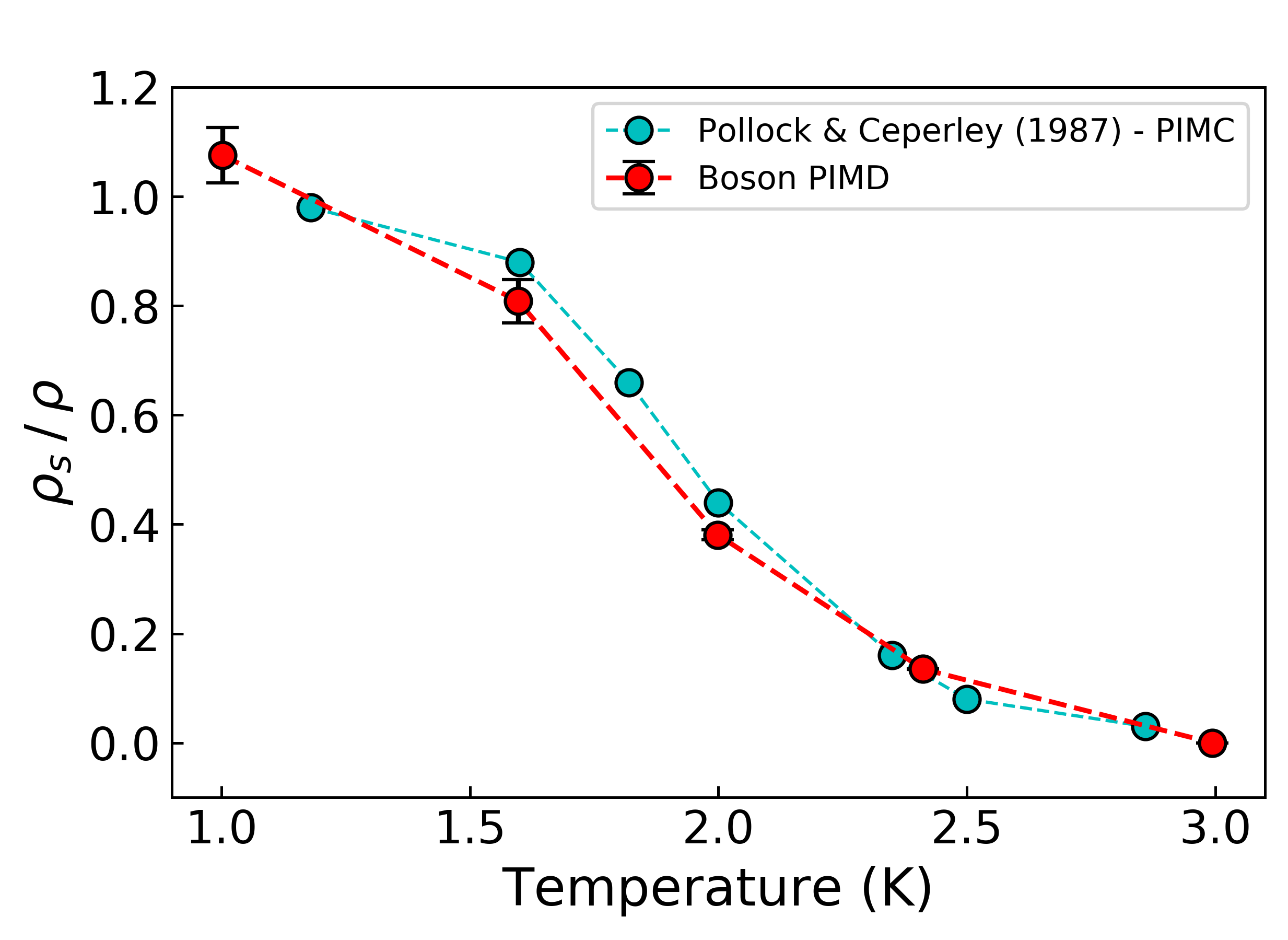} 
\caption{\textbf{A comparison of the superfluid fractions of $^4\text{He}$ liquid superfluid between the PIMC and PIMD-B simulations.} The superfluid fractions $\rho_s/\rho$ as the function of temperatures ranging from 1.0 to 3.0 K calculated by the PIMC (cyan)\cite{Pollock1987} and PIMD-B (red) simulations.}
\label{fig:rhos_pimc_pimd}
\end{figure}

\begin{figure}[h]
\includegraphics[width=15.5cm]{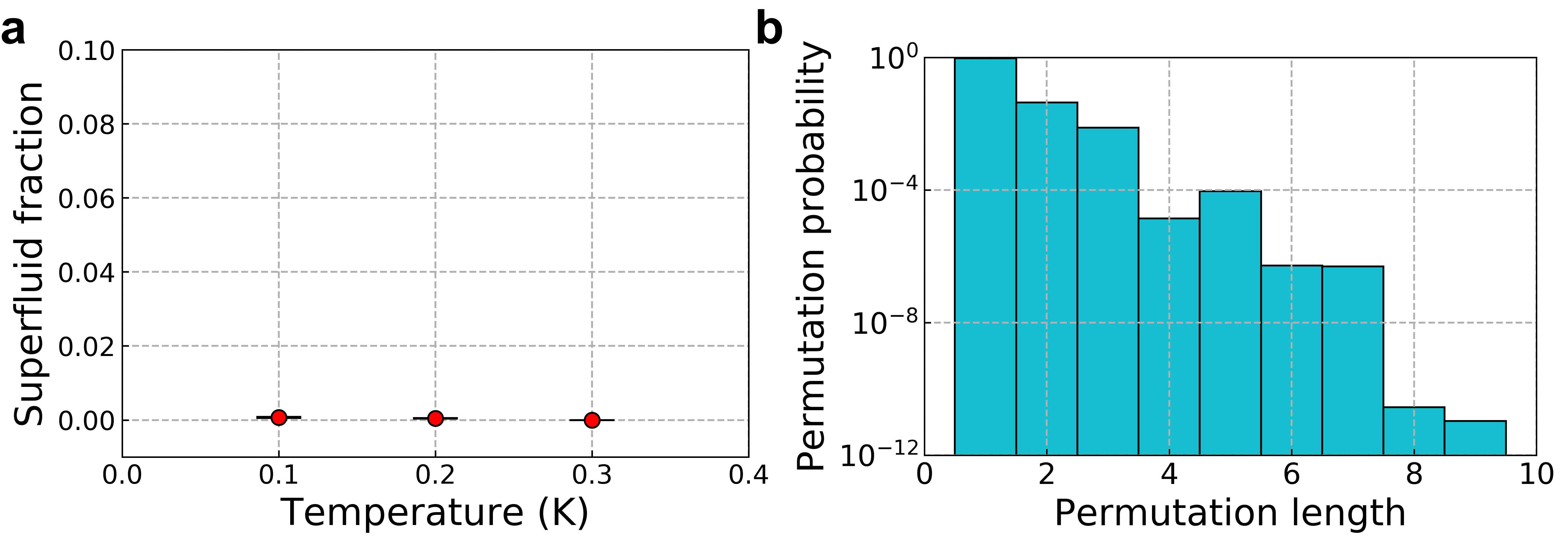} 
\caption{\textbf{Benchmark PIMD-B simulation result of hcp He solid.} (a) The superfluid fractions $\rho_s/\rho$ as the function of temperatures ranging from 0.1 to 0.3 K calculated by the PIMD-B (red) simulations. (b) The permutation probabilities of length $l$ of $hcp$ He solid at $p = 5.5$ MPa and T = 0.1 K.}
\label{fig:rhos_he_solid}
\end{figure}

To validate our approach, we performed a benchmark on the superfluid liquid $^4\text{He}$ system using the \texttt{HFDHE2} potential\cite{Aziz1977} (Supplemental Fig. \ref{fig:rhos_pimc_pimd}). \ctext[RGB]{255,251,204}{We also did a benchmark on the $hcp$ $^4\text{He}$ solid at $p$ = 5.5 MPa between T = 0.1 K and T = 0.3 K using $P = 64$ beads. To eliminate any finite size effects in PIMD-B simulation, we set a sufficiently large unit cell with 216 atoms}\cite{Ceperley04}. \ctext[RGB]{255,251,204}{Although the experiment reported a superfluid transition around T $\sim$ 0.2 K}\cite{Kim2004}, \ctext[RGB]{255,251,204}{the PIMC simulation observed no sign of superfluid transition}\cite{Ceperley04}. \ctext[RGB]{255,251,204}{As expected, our PIMD-B simulation also do not observe the superfluid transition below T $\sim$ 0.2 K (Supplemental Fig.} \ref{fig:rhos_he_solid} \ctext[RGB]{255,251,204}{a). The permutation probability decays exponentially even below the transition temperature (T = 0.1 K), allowing only local permutations (Supplemental Fig.} \ref{fig:rhos_he_solid}\ctext[RGB]{255,251,204}{b)}\cite{Ceperley04}.

\subsection{Permutation probability in PIMD-B simulation}

\begin{figure*}[]
\centering
\includegraphics[width=16.5cm]{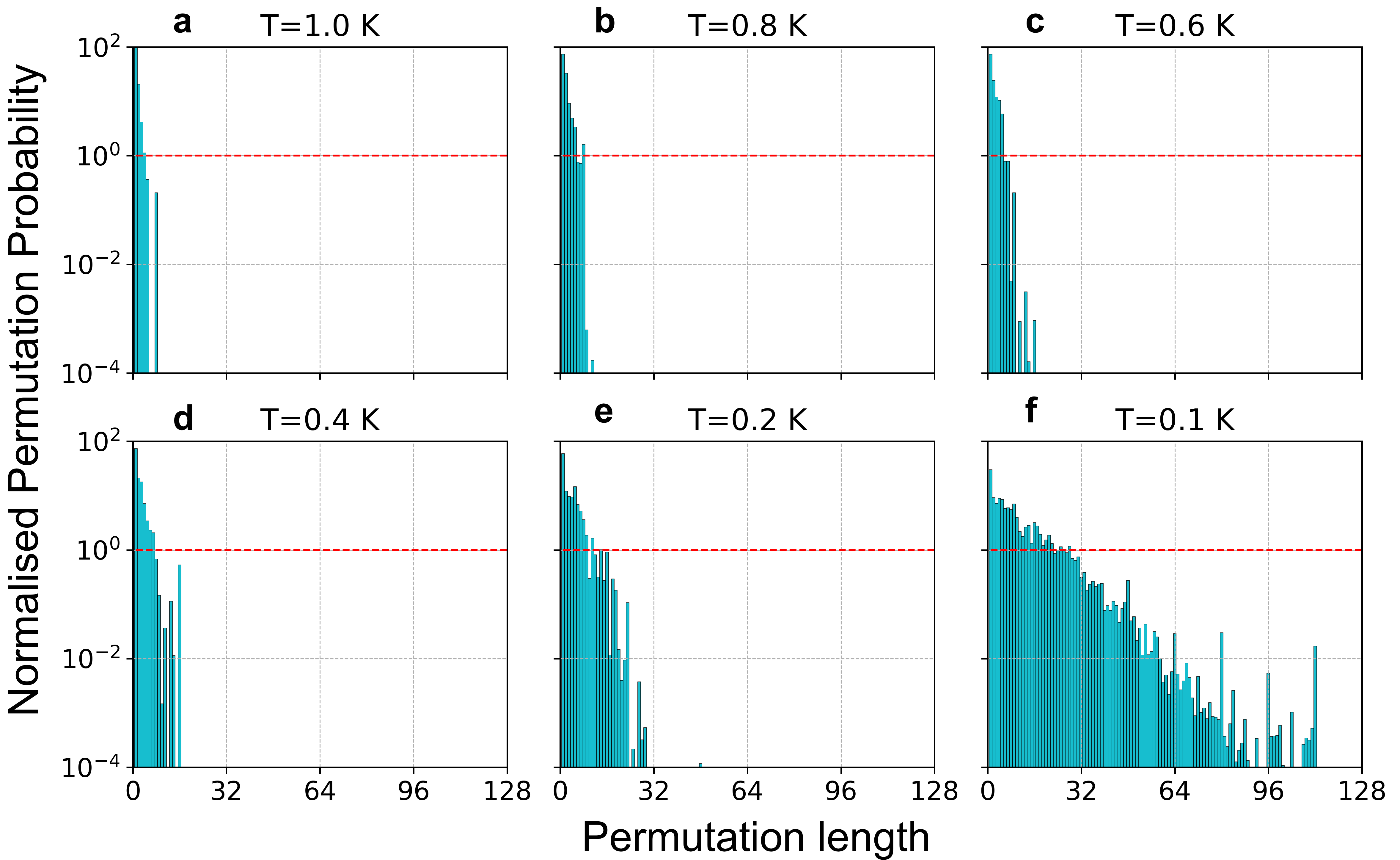}
\caption{\textbf{Normalised permutation probability of high-pressure deuterium.} Normalised permutation probabilities of length $l$ of high-pressure deuterium at $p=800$ GPa are plotted in cyan bar as a function of temperature (a) T = 1.0 K, (b) T = 0.8 K, (c) T = 0.6 K, (d) T = 0.4 K, (e) T = 0.2 K and (f) T = 0.1 K. The permutation probability is normalised by $1/N$. The red dashed lines indicate the equal permutation probability of any $l$ permutation, which is 1.0 in the normalised probability.}
\label{fig:permutation_prob}
\end{figure*}

\floatsetup[figure]{style=plain,subcapbesideposition=top}
\begin{figure}[h]
\includegraphics[height=9cm]{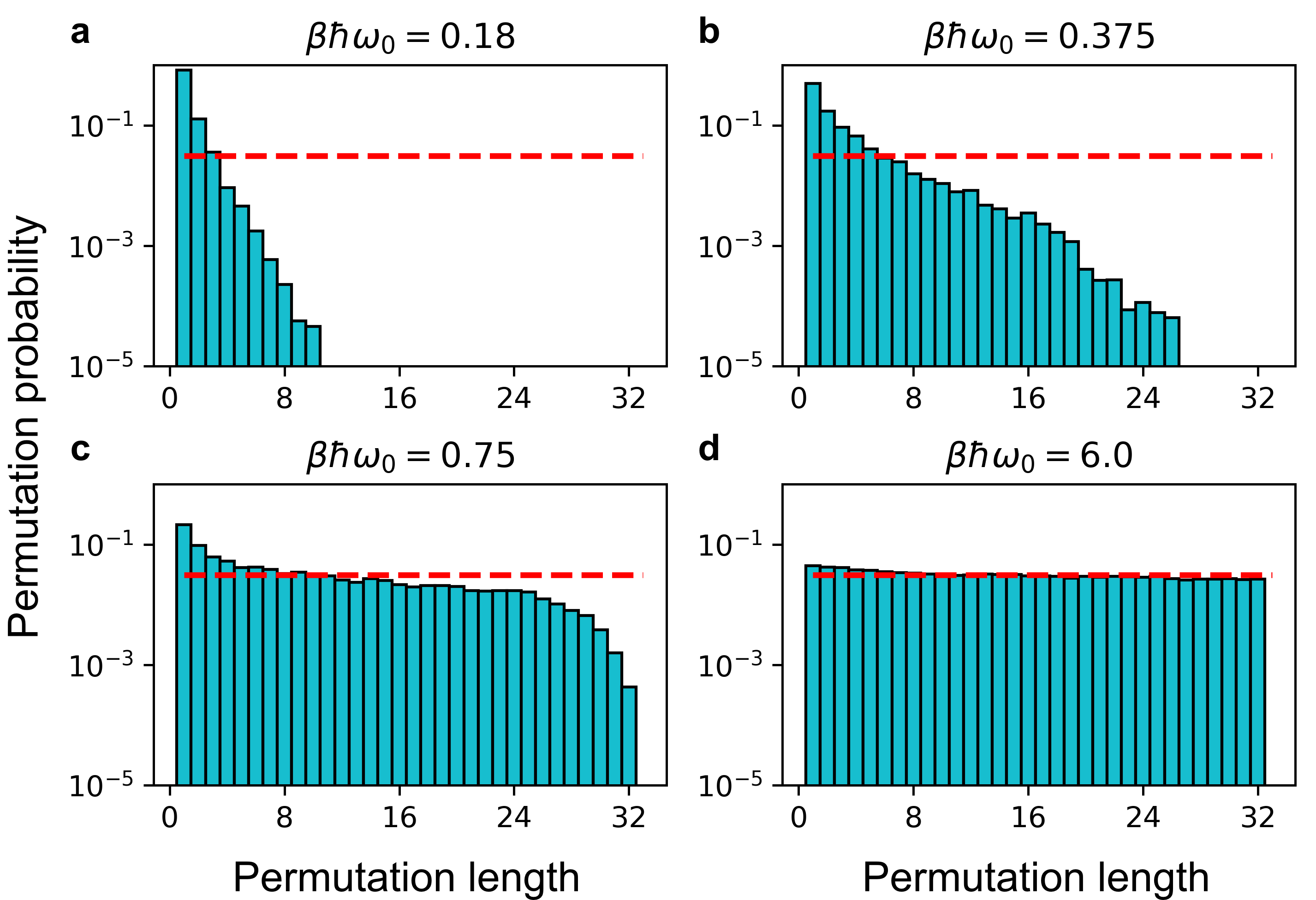} 
\caption{\textbf{Permutation probability of trapped Bosons in a 2D potential.} Permutation probability distributions of $l$ Bosons $p_l$ are plotted as cyan bars with respect to the temperatures (a) $\beta \hbar \omega_0=0.18$, (b) $\beta \hbar \omega_0=0.375$, (c) $\beta \hbar \omega_0=0.75$ and (d) $\beta \hbar \omega_0=6.0$. The red dashed line at each panel is $1/N$ where $N=32$.}
\label{fig:2dboson}
\end{figure}

Following a procedure already established in ref. \cite{Hirshberg2019}, we measure the probability of observing a permutation involving $l$ particle that can be directly extracted from equation (\ref{eq:pl}) (Supplemental Fig. \ref{fig:permutation_prob} and \ref{fig:2dboson}),

\begin{eqnarray}
\label{eq:pl}
p_l =  \frac{e^{-\beta (E^{(l)}_N+V_N^{(N-l)})}}{\sum^N_{k=1} e^{-\beta (E^{(k)}_N+V_N^{(N-k)})}},
\end{eqnarray}
where $V^{(N-l)}_N$ is the bosonic potential of $(N-l)$ particles and $E^{(l)}_N$ is the spring energy of all the beads of $l$ particles. It has been shown that in the limit of a perfect superfluid, this probability is constant as a function of $l$ and equal to $1/N$. Thus a uniform probability can be understood as a sign of superfluidity\cite{Ceperley1995,Krauth96}. 

The equation (\ref{eq:pl}) measures the probability of $l$ permutation occurrence, $p_l$. The most dominant term in the numerator $e^{-\beta(E^{(l)}_N+V_N^{(N-l)})}$ is the potential of the configuration composed of a ring of exchanged $l$ particles and $(N-l)$ independent rings. When normalised by $\sum^N_{k=1} e^{-\beta ( E^{(k)}_N+V_N^{(N-k)})}$, the probability of the remaining configurations are vanishingly small. To this effect, we are able to measure the permutation probability of $l$ bosonic particles. 

As shown in Supplemental Fig. \ref{fig:permutation_prob}, as the temperature of solid deuterium is lowered from 1.0 K to 0.1 K at $p=800$ GPa, the probability of observing long paths increases significantly. At the low temperature limit, the permutation probability approaches to the Bose-Einstein condensation (Supplemental Fig. \ref{fig:permutation_prob}f).

To validate our approach, we perform the benchmark calculations of trapped bosonic particles in a 2D harmonic potential\cite{Hirshberg2019}. We observe the permutation probability of 32 non-interacting bosonic particles at various temperatures $\beta \hbar \omega_0=0.18, \ 0.375, \ 0.75, \ 6.0$, where the 2D trap frequency is $\hbar \omega_0=3$ meV and $\beta = 1/k_BT$ (Supplemental Fig.   \ref{fig:2dboson}). It is well-known that the density matrix elements of any permutations $l$ are only dependent on the ground state in Bose-Einstein condensation\cite{Ceperley1995,Krauth96}. At the zero temperature limit $\beta \hbar \omega_0 \to \infty$ (Supplemental Fig. \ref{fig:2dboson}d), the permutation probability $p_l$ becomes equally probable $1/N$ where $N=32$. Also, the permutation probability recovers the distinguishable particle behaviour at the high temperature limit $\beta \hbar \omega_0 \to 0$ (Supplemental Fig. \ref{fig:2dboson}a).

\subsection{Density of states and inverse participation ratio of a supersolid phase}

We calculate the density of states of a supersolid phase by averaging over $P=256$ imaginary time slices. Since the density of states at given real time steps are similar, we sampled $\sim 10$ configurations at every $1 ps$ and averaged them all. The result (Supplemental Fig.   2d of main text) indicates that the deuterium supersolid is metallic under significant exchange effects. 

To quantify the localisation properties of electronic states of supersolid phase, we calculate the inverse participation ratio (IPR) $p^{-1}_n$ of a given electronic eigenstate $u_n$ defined as

\begin{eqnarray}
\nonumber
p^{-1}_n = & \frac{\sum^N_i |\phi_{n,i}|^4}{\{ \sum^N_i |\phi_{n,i}|^2 \}^2}\\
         = & \sum^N_i |q_{n,i}|^2
\end{eqnarray}
where $\phi_{n,i}$ is a projected atomic wavefunction of eigenstate $u_n$ to the atomic site of atom $i$, $q_{n,i}$ is a projected Löwdin charge population and $N$ is the number of atoms. IPR is a useful measure of localisation of any quantum states. If any states are localised at a particular atomic site $i$, $u_n \sim \delta(\mathbf{r}-\mathbf{r}_i)$, $p^{-1}_n$ becomes unity. On the other hands, $p^{-1}_n$ is $1/N$ for a perfectly delocalised quantum state. 

\subsection{PIMD-B two-phase simulation}

\floatsetup[figure]{style=plain,subcapbesideposition=top}
\begin{figure}[h]
\includegraphics[width=16cm]{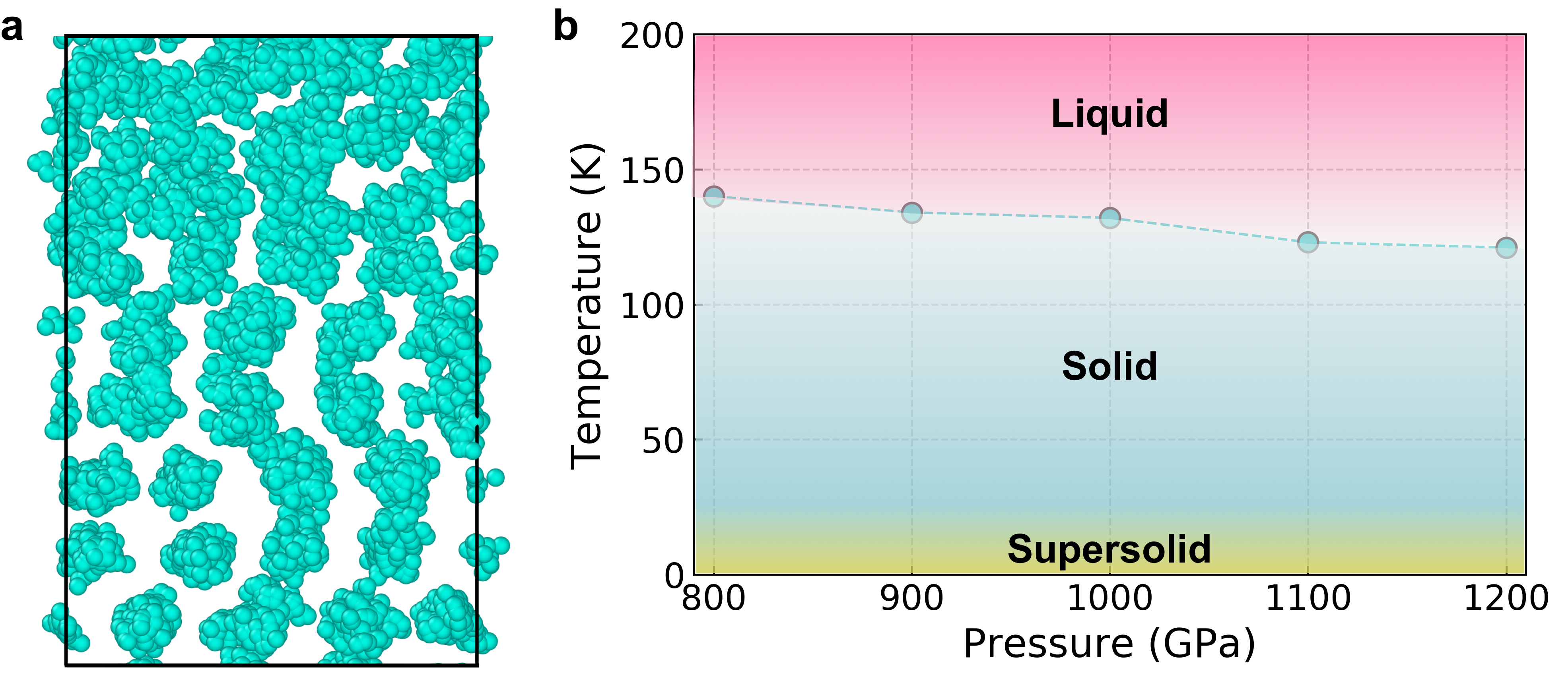} 
\caption{\textbf{Two-phase simulation and melting-line of high-pressure deuterium.} (a) A snapshot taken from two-phase PIMD-B simulation of solid-liquid deuterium interface. (b) P-T phase diagram of high-pressure deuterium at the ranges of $800$ GPa $<p<1200$ GPa and at $50$ K $<T<150$ K.}
\label{fig:2ph}
\end{figure}


The melting points of high-pressure deuterium are estimated at the pressure range of $800-1200$ GPa using ML potential and PIMD-B simulation. We prepared an initial configuration of the interface between the $I4_1/amd$ solid (128 atoms) and liquid (128 atoms) phases, following the previous DFT PIMD simulations\cite{chen13} (Supplemental Fig.   \ref{fig:2ph}a). The initial configuration becomes either solid or liquid depending on the target temperature of PIMD-B simulation. We used $P=64$ beads at which previous DFT PIMD simulation converged\cite{chen13}. Although the DFT PIMD two-phase simulation of hydrogen suggested the existence of liquid metallic ground state from the negative slope of melting curve ($dP/dT<0$), we observe only a slight decrease of melting curve for deuterium. 

\clearpage

\bibliography{supplement}